\documentclass[twocolumn,aa]{aa}
\usepackage[varg]{txfonts}
\usepackage{xcolor}
\usepackage{placeins}
\usepackage{graphicx}

\usepackage{xcolor}

\usepackage{soul}
\usepackage{amsmath,amssymb}
\usepackage[normalem]{ulem}
\usepackage{color,booktabs}
\usepackage[toc,page]{appendix}

\begin{document}
\title{Identifying preflare spectral features using explainable artificial intelligence}
\author{Brandon Panos\inst{1,2} \and Lucia Kleint\inst{1,2} \and Jonas Zbinden\inst{1,2}}
\institute{University of Geneva, 7, route de Drize, 1227 Carouge, Switzerland
\and
Astronomical Institute of the University of Bern, Sidlerstrasse 5, 3012 Bern}
\date{Received 29 August; Accepted 24 December 2022}

\abstract{
The prediction of solar flares is of practical and scientific interest; however, many machine learning methods used for this prediction task do not provide the physical explanations behind a model's performance. We made use of two recently developed explainable artificial intelligence techniques called gradient-weighted class activation mapping (Grad-CAM) and expected gradients (EG) to reveal the decision-making process behind a high-performance neural network that has been trained to distinguish between \ion{Mg}{II} spectra derived from flaring and nonflaring active regions, a fact that can be applied to the task of short timescale flare forecasting. The two techniques generate visual explanations (heatmaps) that can be projected back onto the spectra, allowing for the identification of features that are strongly associated with precursory flare activity. We automated the search for explainable interpretations on the level of individual wavelengths, and provide multiple examples of flare prediction using IRIS spectral data, finding that prediction scores in general increase before flare onset. Large IRIS rasters that cover a significant portion of the active region and coincide with small preflare brightenings both in IRIS and SDO/AIA images tend to lead to better forecasts. The models reveal that \ion{Mg}{ii} triplet emission, flows, as well as broad and highly asymmetric spectra are all important for the task of flare prediction. Additionally, we find that intensity is only weakly correlated to a spectrum's prediction score, meaning that low intensity spectra can still be of great importance for the flare prediction task, and that $78$\% of the time, the position of the model's maximum attention along the slit during the preflare phase is predictive of the location of the flare's maximum UV emission.}

\maketitle

\section{Introduction}
A solar flare is a sudden release of energy due to magnetic reconnection, resulting in an enhancement across the entire electromagnetic spectrum \citep{Carmichael_1964,Sturrock_1966,Hirayama_1974,Kopp_1976,Adding_free_energy}. For large flares this energy can reach an order of $\sim10^{32}$ ergs \citep{Aulanier_2013}, and is used to accelerate charged particles toward and away from the solar surface. The energy dissipates over stages with timescales that vary from seconds to hours \citep[e.g.,][]{Fletcher2011}. The outward bound material fills the interplanetary medium with high-energy particles and perturbs the heliospheric magnetic field \citep{Rouillard_2016}, both of which interact with Earth's magnetosphere. These events, if directed toward Earth, can trigger power grid blackouts and adversely affect communication systems \citep[e.g.,][]{boteler2006, Schrijver2014}, which in today's technologically saturated environment comes with a high socioeconomic cost, making their prediction critical.\\

Machine learning has provided us with a set of powerful algorithms that can be used to attempt to predict whether an active region (a patch on the Sun associated with enhanced magnetic activity), will produce a flare or not. One of the most successful algorithms for such a task is the so-called neural network (NN) \citep{Rosenblatt_1958}, which is a computational graph-like structure that draws inspiration directly from the mammalian brain. Just as an organic brain, these networks automatically program themselves through experience \citep[e.g.,][]{Ian_2016} and can generalize what they have learned to make informed decisions about new observations \citep[e.g.,][]{Vidyasagar_2003}.

The majority of flare prediction efforts rely on full photospheric vector magnetograms recorded by the Helioseismic and Magnetic Imager \citep[HMI,][]{Scherrer_2012,Hoeksema_2014} onboard the Solar Dynamic Observatory \citep[SDO,][]{Lemen_2012}. These photospheric magnetic data are typically fed into a machine learning algorithm such as a NN \citep[e.g.,][]{bobra2015solar,florios2018forecasting,Liu_2019,So_s_2022}. In terms of the true skill statistic (TSS), the standard metric for evaluating a model's predictive performance (1 being optimal and -1 being adverse), the expected baseline TSS using data from SDO is around $\sim 0.7$, regardless of the sophistication of the machine learning algorithm \citep{Monica_Corona_HMI}. This indicates an apparent bottleneck and limit to the utility of photospheric magnetic data.

As a way to overcome this bottleneck, researchers have started experimenting with novel parameterizations of the HMI magnetic data using topological data analysis to codify their rich spatial features \citep{Deshmukh2020, Deshmukh2022}. Moreover, additional information from complementary data sources such as soft X-ray, flare history, and AIA photospheric, chromospheric, and coronal images have been incorporated in an attempt to improve model performance \citep{UV_Brightening,Monica_Corona_HMI}.

New evidence suggests that high resolution spectral data can be used to predict solar flares at least on short subhour timescales. \citet{Panos2020} created two classes of \ion{Mg}{II} spectra captured by NASA's Interface Region Imaging Spectrograph (IRIS) \citep{IRIS}. The first class consisted of spectra from active regions that did not lead to solar flares (hereafter referred to as the AR class), and the second one consisted of spectra collected from active regions $25$ minutes before flare onset (referred to as the preflare PF class). It was demonstrated that a simple feed forward fully connected NN could distinguish between spectra from either class with an $80\%$ accuracy, precision, and recall. Furthermore, the network's performance increased monotonically when successively feeding spectra from $t=30$ to $t=0$ minutes before the onset of a large X1.6-class flare, albeit over a restricted region of the slit. It is not, however, clear why the network performed so well at the classification task, and on what grounds it based its decisions. Nevertheless, these results open up the possibility of not only improving the current performance of our models, but due to the high diagnostic capabilities of spectra \citep[e.g.,][]{Leenaarts_2013A}, could also provide critical information about the state of the preflare atmosphere, and any necessary conditions that might facilitate a solar flare \citep[e.g.,][]{yang_2019}, thus shifting the focus from model performance to physical understanding.

Along these lines, and in an attempt to explain the above results, a recent study made use of a clustering technique to identify common spectroscopic precursors and atmospheric conditions that might facilitate flare triggering events \citep{Woods_2021}. It was found that single-peak emission in both the \ion{Mg}{II} h\&k lines as well as the pair of subordinate lines located at $\sim2798.8~\text{\AA}$ appeared most commonly, but not exclusively, within the study's PF dataset. Inversions of these single-peaked profiles using the STiC inversion code \citep{Cruz_2019A} indicate enhanced chromospheric temperatures and electron densities. Similar to previous findings \citep[e.g.,][]{Cheng_1984,Machado_1988,Harra_2001,Panos2020}, the authors speculate this to be a consequence of small-scale heating events possibly driven by reconnection as far back as $40$ minutes before flare onset. Since this clustering approach of identifying important PF spectra is manually intensive and time-consuming, a recent study automated the process via the use of multiple instance learning (MIL) \citep{Huwyler_2022}. In addition to high accuracies on the AR/PF classification task, their models automatically identified spectra that were judged to be important for flare prediction, confirming the results found in \citet{Woods_2021}. Their work however does not indicate the particular features of each spectrum that are responsible for high model scores.\\

For this study, we made use of the same \ion{Mg}{II} dataset from the original paper. We then trained a powerful "visual" NN called a convolutional neural network or ConvNet on the AR/PF binary classification problem. Once the model learned to distinguish between AR and PF spectra, we used a class of techniques, collectively referred to as Explainable Artificial Intelligence (XAI) \citep{Barredo_2019} to derive direct explanations from the ConvNet without having to perform intermediate manual steps. These techniques allowed us not only to automatically discover which spectra are important, but which features of the spectra are most critical for predicting solar flares.

\begin{figure*}[t]
\centering
\includegraphics[width=0.9\textwidth]{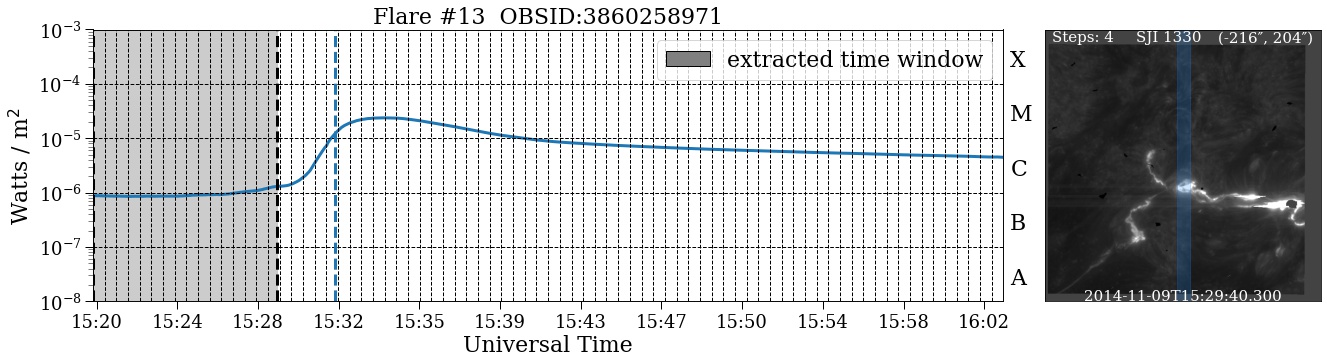}
\includegraphics[width=0.9\textwidth]{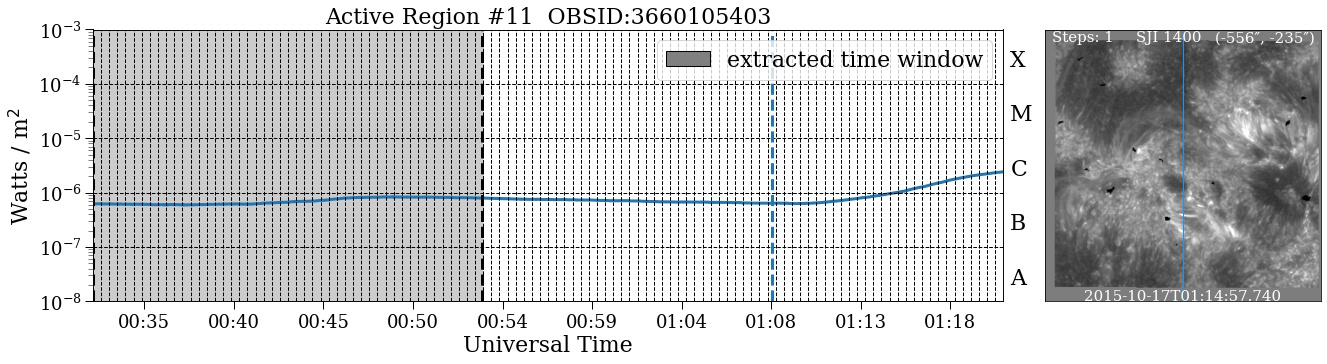}
\label{overview}
\caption{Example of data collection for a single PF (top) and AR (bottom) observation. \ion{Mg}{II} spectra in both cases were extracted from the gray time interval with the interval between each vertical dashed line indicating the completion of a single raster. For context a SJI is included for each observation at the time roughly indicated by the vertical dashed blue line. The blue curves in the time plots indicate the integrated GOES flux in the $1-8~\text{\AA}$ channel, while the blue regions in the SJI indicate the span of the raster, with a four step raster for the PF observation and a sit-and-stare for the AR observation.}
\end{figure*}

\section{Data}
\label{data_section}
The data used in this study are composed of \ion{Mg}{II} spectra captured by the IRIS satellite. For consistency, the dataset is identical to that used in \citet{Panos2020}. As before, we selected a spectral window spanning $2794.14-2805.72~\text{\AA}$, which includes both the \ion{Mg}{II} h\&k lines as well as the triplet emission around $2798.77~\text{\AA}$. We then partitioned observations into two classes based on the GOES soft X-ray flux: The first partition is called the AR class, and is composed of roughly $2.5$ million spectra derived from $18$ observations and extracted from a $25$ minute window at the start of each observation. An observation here means a period with a predefined observing scheme, which can last several hours. The only criterion for an observation in this class was that a large flare (M- or X-class) did not appear over the entire duration of the IRIS observation, as verified by the GOES X-ray flux. This does not preclude smaller flares such as C-class and lower. 
Since this study probes short timescales (as apposed to the normal $24$ hour margins used in flare prediction studies), it is not guaranteed that the active region did not produce a large flare after the IRIS observation's time window ended, leading to a possible weak mixing of the classes.

The second partition is called the PF class and consists of $2.4$ million spectra taken from $19$ observations in a time window $25$ minutes (sometimes shorter if IRIS was not recording) before each X- or M-class flare's onset as defined by the NOAA catalog flare start time (Table $1$ of the same study). 
Like previous studies \citep[e.g.,][]{Monica_Corona_HMI,Angryk_2020}, we do not consider small flares ($<$M-class) as targets for prediction, but rather as a set of important features whose frequency could help predict larger events, meaning that smaller flares were not necessarily precluded from the $25$ minute analysis window for both the AR- and PF class.

An example of a PF (top) and AR (bottom) observation can be seen in Fig.~\ref{overview}. The gray regions indicate the $25$ minute time windows over which \ion{Mg}{II} spectra were collected with the time in minutes on the x-axis. As indicated by the GOES $1-8~\text{\AA}$ flux (solid blue curve) and flare class along the y-axis, a large M-class flare occurs in the PF observation directly after the sampling region, while the AR observation is flat and then slightly raises to a C-class level towards the end of the observation. It is important to remember that GOES integrates the X-ray flux from the entire solar disk while IRIS observes only a small patch of the sun (a maximum of $175\times175~\text{arcsec}^2$ for the slit-jaw imager and $130\times175~\text{arcsec}^2$ for its spectrometer), therefore, although the large flares were visually confirmed to be within IRIS's field of view (and indeed at some future point crossing the spectrogram), the X-ray signals in the AR class could have occurred somewhere else on the solar disk. Each interval between the vertical black dotted lines indicates the completion of a single IRIS raster. If the observation is a sit-and-stare, meaning no spatial rastering of the spectrogram, as in the AR example, then the intervals indicate the individual exposure times. For context, a SJI accompanies each figure and was rendered roughly around the time indicated by the vertical blue dashed lines. The blue regions within each SJI indicate the spatial extent of the IRIS raster which is spanned by steps of equal size. Each flare was assigned a number in the title for future reference.

After sampling the spectra using the above prescriptions, a series of additional processing steps were performed on the already preprocessed IRIS level 2 data to make it amenable to machine learning methods. These steps were as follows: 1) All spectra were interpolated to a grid size of $240~ \lambda$-points. 2) Each spectral profile was normalized by its maximum value, effectively placing precedence on the shape of the \ion{Mg}{II} spectra instead of their intensity. It is important to keep in mind that the intensity is still weakly encoded into the spectral shape, since high intensity profiles, once normalized, will result in spectra with flat continuum emission. To avoid noisy and corrupt data, entire spectra were replaced with nan's if they met any of the following conditions: Missing data $\to$ If a spectrum (within our selected wavelength window) contained at least one value $<-100$. Overexposure $\to$ If $10$ consecutive intensity values were equal to the maximum intensity value. This is quite a high threshold and in retrospect should be decreased, however, for the sake of continuity with previous results we maintain this setting. Poor signal-to-noise $\to$ Any spectra that remained below a 10 $\text{DN/s}$ threshold. Cosmic rays $\to$ If the maximum intensity value appeared outside of a small window surrounding both the h\&k-cores, or if the k/h-ratio exceeded the theoretical limits of 2:1 or 1:1 by a value of $0.3$, that is spectra with ratios exceeding $2.3$ or less than $0.7$ were excluded. Additionally, the reconstruction error of a variational autoencoder trained on quiet Sun spectra was used as a way to dynamically mask spectra that are typically associated with quiescent Sun observations, see \citet{Panos_2021} for details. 

\section{Model development}
In this section, we build up the intuition necessary to understand how our machine learning algorithm predicts whether a spectrum comes from a nonflaring or flaring active region by introducing a general set of machine learning principles and training etiquette. We then discuss our particular case, including a description of the model that we use (a ConvNet), and how IRIS data in combination with the search for quality explanations in addition to high performance, forces us to modify the standard training practice.

\subsection{Problem outline}
\label{general_procedure}
Our goal is to create a function/model $\mathcal{F}_\Theta(x)\to \hat{y}$, that is parameterized by a set of arguments $\{\theta~|~\theta\in\Theta\}$, and can take as input a raw spectrum $x\in X$ from the dataset, and produce an output probability $\hat{y}\in[0,1]$ representing the model's guess as to whether the spectrum belongs to either the AR (output closer to zero) or PF class (output closer to one). The objective is to find a set of parameters $\{\theta\}$, referred to as the model's weights, that generates the closest match between the actual set of labels $\{y_n\}$ and the predicted set of labels $\{\hat{y}_n\}$, for example, we want a model that can correctly predict the class of as many spectra as possible. This type of problem is referred to as a binary classification problem and can be solved using any number of methods. The optimal parameterization is usually found by introducing a loss function $\mathcal{L}(y, \hat{y}, \Theta)$, that when minimized, guides the model parameters in a controlled manner to their optimal values $\{\theta^*\}$. This process of incrementally adjusting the model's weights is referred to as training and is achieved via gradient descent, that is, by taking the derivative of the loss function $\partial_\theta \mathcal{L(\theta)}$ with respect to the weights, and adjusting those parameters in small increments in the direction that results in the largest decrease to the loss. Since the objective is simply to have a model make as many correct label predictions as possible, the loss function depends on the true and predicted labels as well as its internal set of parameters. To train our models we made use of the binary cross-entropy (BCE) loss function (with regularization)
\begin{equation}
\begin{split}
\mathcal{L}(y, \hat{y}, \Theta) &= -\frac{1}{N} \sum_{n=1}^{N}\Big\{y_{n} \log \mathcal{F}_\Theta(x_n) \\
&+\left(1-y_{n}\right) \log \left[1-\mathcal{F}_\Theta(x_n)\right]\Big\} \\
& + \lambda \sum^M_{m=1}\theta^2_m,
\end{split}
\label{BCE}
\end{equation}
which is a measure of dissimilarity between the true set of labels $\{y_n\}$ and the predicted set of labels $\{\hat{y}_n\}$. If the differences between these two sets is large, then the loss is also large, and vice versa. In terms of information theory, the BCE is the average number of bits required to decode a signal using an estimated distribution rather than the true distribution. It is also equivalent to maximizing the log-likelihood of the data under $\{\theta\}$ and is identical to the KL-divergence up to an additive constant, where the KL-divergence is a pseudo metric used to measure the dissimilarity/distance between two probability distributions.

With large datasets, it is computationally expensive to only update the weights after computing the sum of the loss over every training example (a single epoch). For this reason, we use a variant of gradient descent called stochastic gradient descent \citep{saad_1999}, which updates the weights based on a pseudo loss that is estimated by randomly sampling a subset of spectra instead of all spectra. In our case, $N$ only runs over $64$ spectra for each update.

The last term in Eq.~\ref{BCE} is not part of the definition of the BCE-loss but is added to penalize the weights and help the model generalize to new data. This is necessary since our true objective is not simply to maximize the number of correct predictions with the data the model was trained on, but to develop a robust model that can correctly label new data. The prefactor $\lambda$ is called a regularization parameter and dictates the trade-off between model performance and robustness, with larger values generally leading to poorer performance on the training set but better generalization to new instances. If we do not regularize the model, gradient descent would most likely converge to a set of weights that are overly optimized on the training data. 

To get a true measure of a model’s performance, it is important to test the model with data that it has not encountered during training. It is standard practice to therefore split the data into a training, validation, and test set. While the weights are adjusted using data from the training set, the hyperparameters such as the model architecture as well as the regularization term $\lambda$ are adjusted on the validation set. The test set then serves as the gold standard to judge a model's performance and ability to generalize.

\begin{figure}[thb]
\center
\includegraphics[width=0.49\textwidth]{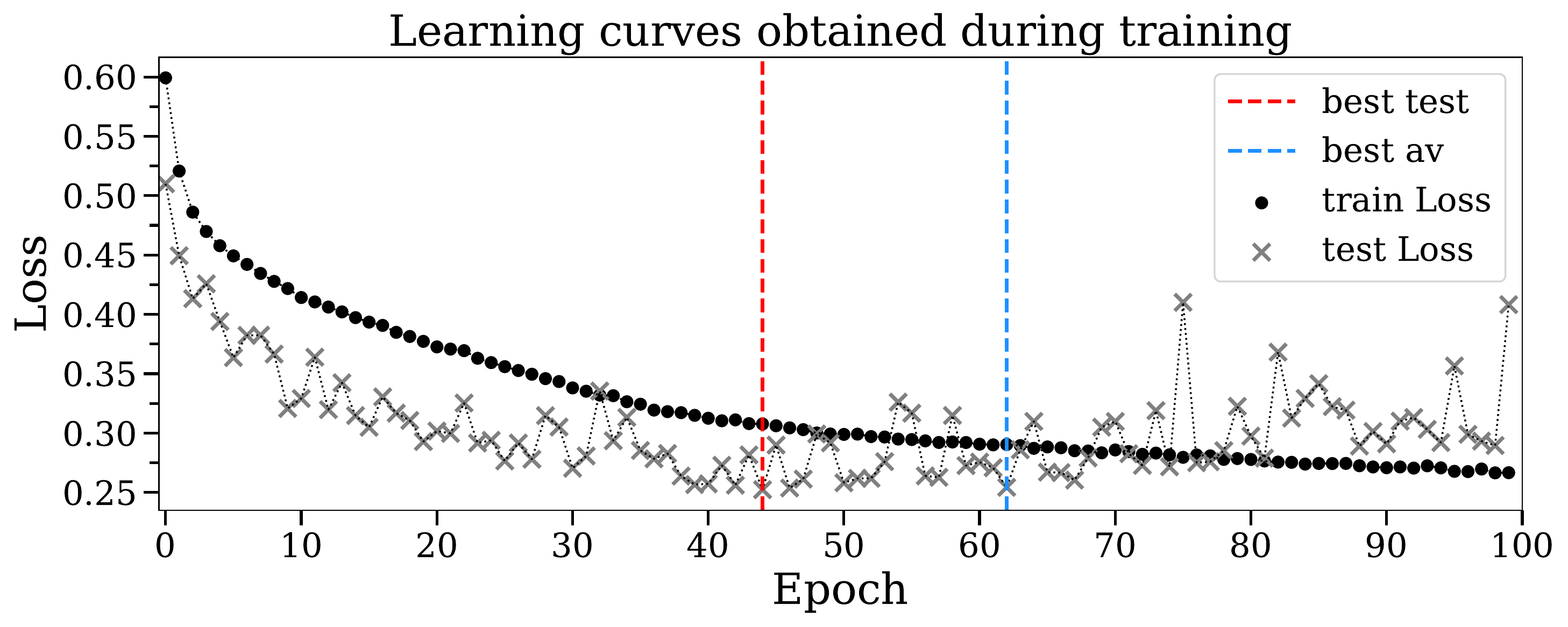} 
\caption{Example of learning curves obtained during training. The black and gray curves indicate the performance (in terms of loss) of a model on both the training (black) and test (gray) set. The best model could either be the model that scored the lowest loss on the test set (red dashed line), or the average lowest loss on both the training and test sets (blue dashed line). Notice that after about 50 epochs the test loss starts to increase, indicating a possible overfitting of the training data.}
\label{learning_curve}
\end{figure}

The quality of the model can then be judged by analyzing the behavior of the loss (or accuracy) as a function of epoch (training time) on both the training and test set. During training, these two losses trace out curves collectively known as learning curves (as seen in Fig.~\ref{learning_curve}), whose joint dynamics can be used to identify unwanted behavior such as under or overfitting the training data, as well as the optimal training time (either red or blue vertical lines). The model with the lowest loss on the test set is usually selected as optimal.\\

\begin{figure*}[thb]
\begin{centering}
\includegraphics[width=0.7\textwidth]{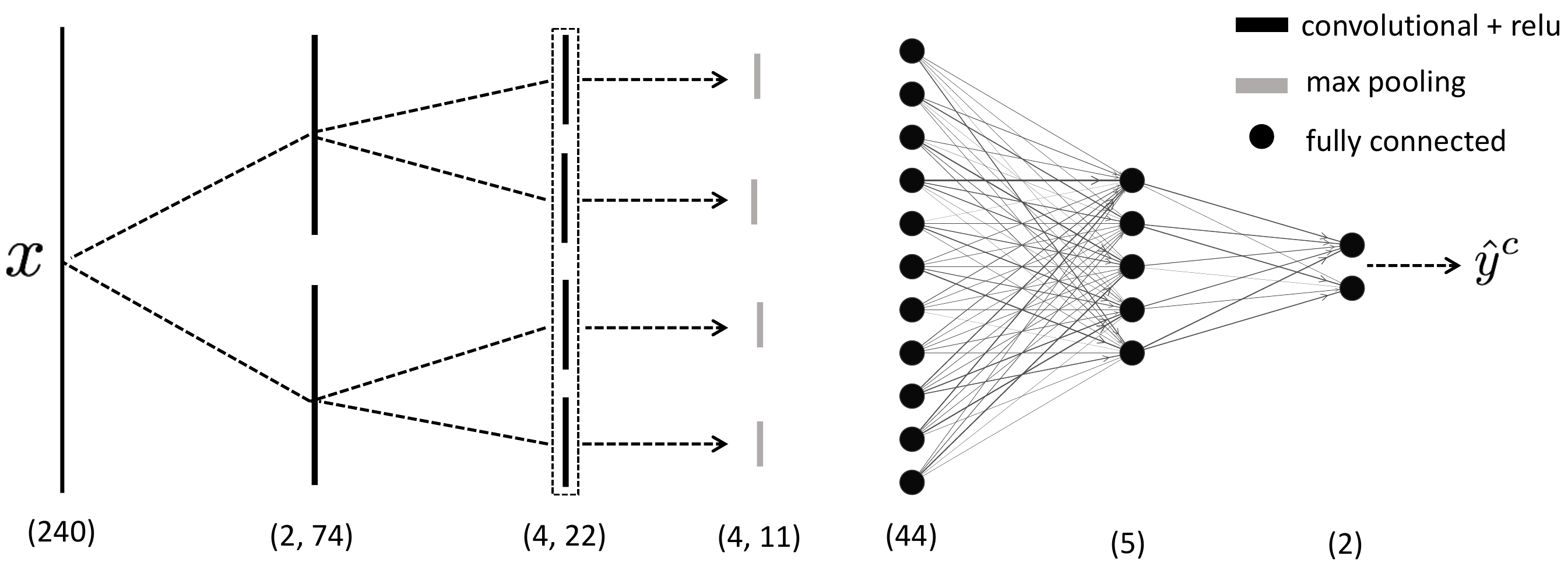} 
\caption{Schematic of a ConvNet. This network is a hybrid between a set of convolutional layers and a fully connected feed forward network. The convolutional layers contain kernels that scan through the input and extract feature maps (vertical black solid lines). Each layer forms successively more abstract and sophisticated maps, with each filter being sensitive to a particular feature that the network deems important for the classification task. The high level features which the network automatically generates are then passed through a max pooling layer (gray vertical lines), which take summary statistics of the maps, reducing the number of parameters and promoting generalization. The final content is then fed into the fully connected layers which behave as described in Fig.~\ref{FFNN}, and are responsible for classifying the spectrum into either the AR or PF class. It is important to note that the weights and kernels are updated during training, meaning that the ConvNet adaptively searches for a set of features that is most helpful for distinguishing between AR and PF spectra.}
\label{ConvNet}
\end{centering}
\end{figure*}

\subsection{Convolutional neural network}
\label{feature_map_section}
Optimization problems such as these are commonly parameterized by NN's, meaning that the function $\mathcal{F}_\Theta$ is replaced by a NN. A ConvNet is typically composed out of two sections as seen in Fig.~\ref{ConvNet}. A feature extraction segment, inspired from the organization of an animal visual cortex \citep{Hubel_1968}, which derives high level interpretable features via the use of stacked convolutional modules \citep[e.g.,][]{Ian_2016}, and the classification segment, composed of multiple fully connected dense layers that are responsible for interpreting those features and mapping them to a binary classification output (AR/PF). For a detailed description of the fully connected component see appendix \ref{Classical_neural_network}. 

\begin{figure}[thb]
\includegraphics[width=0.48\textwidth]{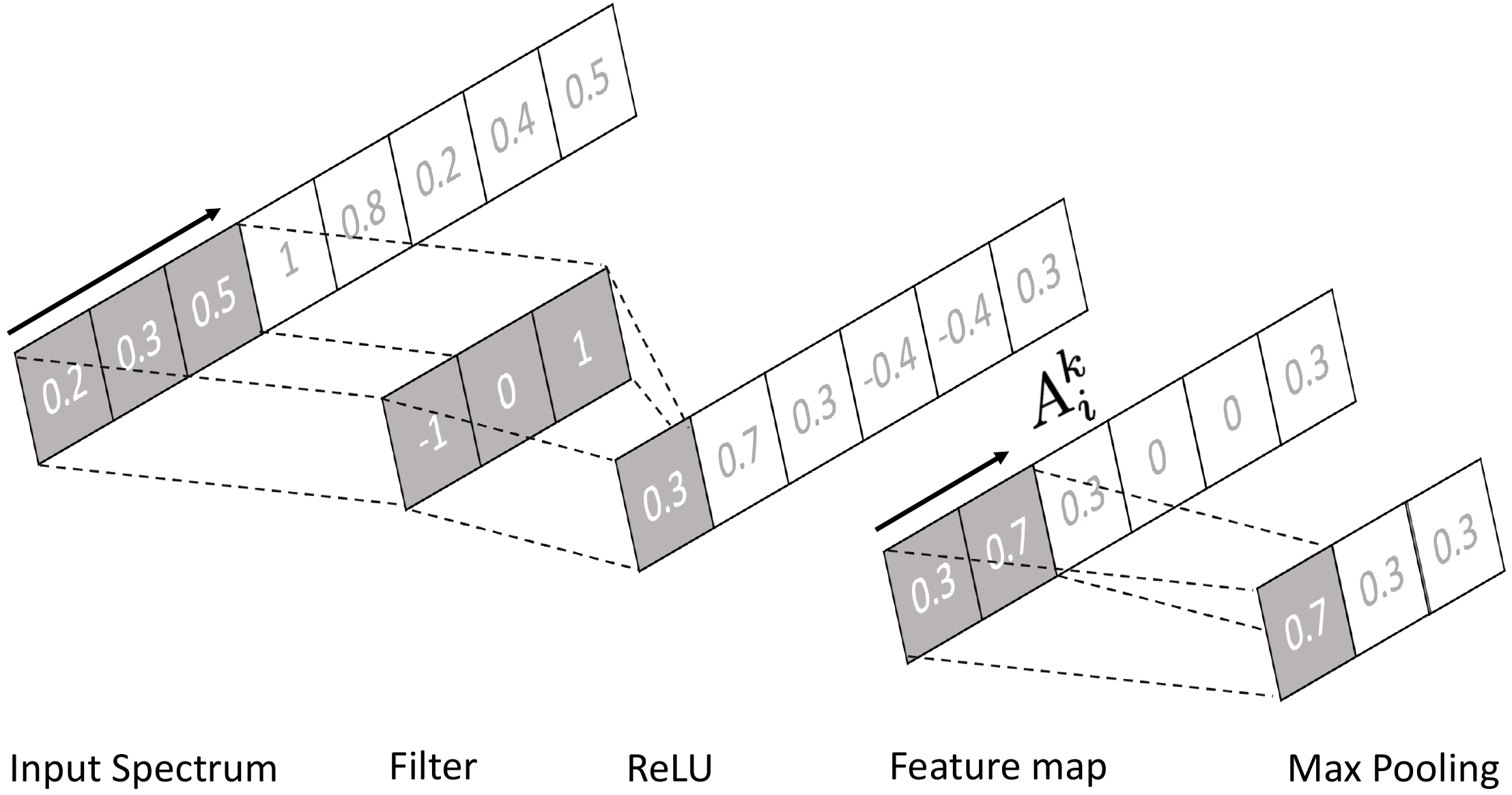} 
\caption{Schematic for the generation of a single feature map. A filter/kernel consisting of a visual field of three pixels scans over the input in strides of one. At each position, the dot-product between the input and kernel is taken to form a single output in the next layer. This process is referred to as a convolution. The resulting output is then passed through a ReLU activation function which introduces a nonlinearity into the network by setting all negative values to zero. The feature map $A^k_i$ can then optionally be passed through a max pooling function, which scans across the map in strides of two and maintains only the largest value. The convolution process makes the network efficient by promoting sparse connections and weight sharing, while the pooling layer further reduces the number of parameters and promotes invariance of the network output to small shifts and rotations of the input.}
\label{convcell}
\end{figure}

ConvNets provide several desirable advantages over simple fully connected networks, most importantly, they provide a means to automatically search for useful features in the input that may help minimize the loss function and lead to better classifications \citep{lecun_1995}. This implies that the convolutional section of the network generates a potential set of "reservoirs" for extracting explanations. Our particular ConvNet contains six such reservoirs referred to as feature maps $A^k$, indicated by black vertical lines in Fig.~\ref{ConvNet}. Each feature map is constructed via a succession of simple layered operations as indicated in Fig.~\ref{convcell}.

Firstly, a filter/ kernel, sweeps across the input with a predefined number of steps between each stride. In this case, the filter has a visual field of three units and a stride of one unit. At each step, the network takes the dot-product of the input and the kernel (a convolution), to produce the set of numbers seen in the next layer. It is important to note that at this stage two critical attributes are injected into the ConvNet that distinguish it from its fully connected counterpart. Firstly, each output node is connected to only three input pixels, unlike a fully connected layer where every neuron is connected to the entire visual field of the input, and secondly, the parameters of the filter (in this example scalar values $[-1,0,1]$) are reused for each convolution across the entire input. This means that not only do we have fewer connections between the layers, but those connections that we do have are restrained to share similar sets of weights $\{\theta\}$, resulting in a drastic improvement in computational efficiency. 

The output vector resulting from these convolutions is then passed through a ReLU activation function which introduces a nonlinearity by replacing any negative numbers by zero. The resulting feature map can optionally be further reduced in size by applying a layer that extracts global statistics, such as a max pooling layer that maintains the largest of two numbers when sweeping over the feature map in strides of two. Pooling layers such as these make the network's output invariant to small shifts or rotations in the input, since only general information across multiple pixels is retained. Additionally, these layers help alleviate overfitting and further reduce the number of parameters fed into the fully connected layers.

Returning back to Fig.~\ref{ConvNet}, each vertical black line represents a feature map that has been obtained using a different filter. The values of each filter also contribute to the set of parameters $\{\theta\}$ that define the network, and are updated in kind via backpropagation (a NN equivalent of gradient descent) in order to minimize the loss function in Eq.~\ref{BCE}. This means that each map tries to identify useful features to serve as an optimal basis for the fully connected layers. 

\begin{figure}[thb]
\center
\includegraphics[width=0.47\textwidth]{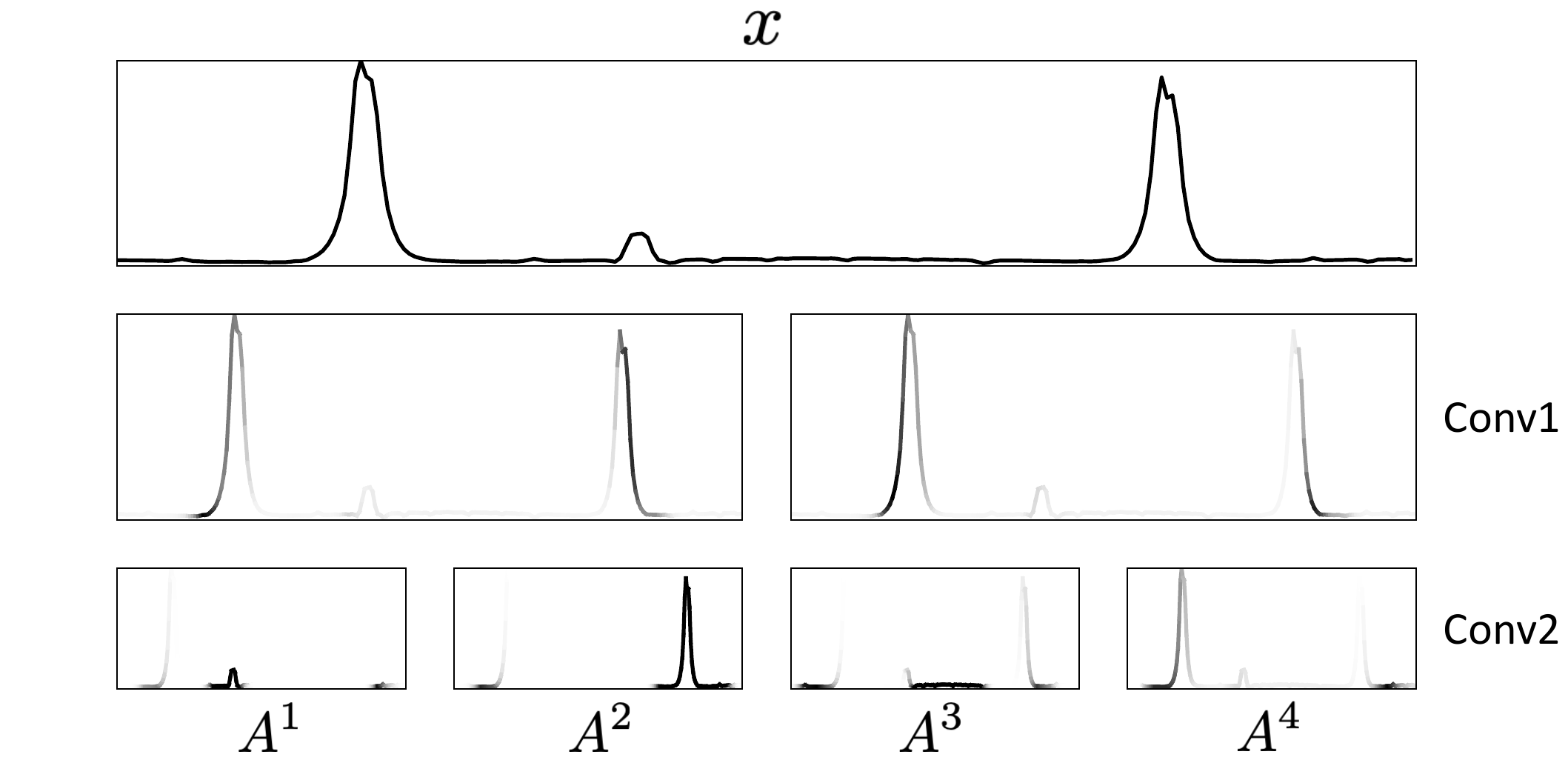} 
\caption{Diagram showing the actual activations of each feature map in response to an input spectrum $x$. These maps correspond to the 6 vertical black solid lines in Fig.~\ref{ConvNet}. Darker shades show what part of the spectrum each map is identifying.}
\label{feature_maps}
\end{figure}

During training, the kernel of each feature map alters its parameters, in order to strengthen responses to particular input patterns. For instance, one of the feature maps could be dedicated to identify triplet emission. If a spectrum has large triplet emission, then this filter will become very active and initiate a strong signal to the fully connected layers. This is demonstrated in Fig.~\ref{feature_maps}, which shows the actual activations of our trained network for each feature map in Fig.~\ref{ConvNet}. In order to see what feature each map is searching for, we have projected the activations back onto the input spectrum, with darker regions indicating importance.

The first convolutional layer contains two maps that differ only slightly from the input. These maps are then convolved in the second layer to produce more abstract and sophisticated features. We can see that in the final layer, the feature map $A^1$ has understood that the input spectrum contains triplet emission, and furthermore, that this is important for the classification task. Map $A^3$ on the other hand is searching for the presence, or lack of pseudo-continuum emission. We therefore see that the convolutional layers act as a type of translator, identifying the most suitable language for the fully connected layers to interpret, and furthermore, that the final convolutional layer contain the most sophisticated understanding of the spectrum.\footnote{The spatial coherency seen in the feature maps is shattered when passed through the fully connected layers.}

The details of our ConvNet can be seen in Fig.~\ref{ConvNet}. The raw spectrum enters into the network, is convolved with two filters that have strides of $3$ and visual fields of 20 pixels each. The convolved outputs are passed through a ReLU activation function to produce two maps of dimension $74$ each (We note that any remaining pixels after taking the modulus of the convolution are dropped). These two maps serve as inputs for the generation of four more abstract maps that were obtained using filters of stride $3$ and visual fields of 10 pixels. The contents of these filters are then reduced via a max pooling layer (stride $2$ kernel size $2$), flattened into a single vector of size $44$, and then fed into the fully connected layers for classification. For reasons discussed in section \ref{GradCAMsection_detailed} of the appendix, this parameterization of the ConvNet represents a minimalistic architecture that proves sufficient for our classification problem.

\section{Explainable AI techniques}
In this section, we briefly describe two different methods, Gradient-weighted Class Activation Mapping (Grad-CAM) and Expected Gradients (EG) for extracting features from \ion{Mg}{II} that are important for flare prediction, while their detailed description can be found in the appendix. Both of these methods are applied to our ConvNet after it has been trained (post-hoc), and produce visual explanations in the form of saliency maps, which are simple heatmaps that can be projected back onto the input spectrum to indicate the most critical features that were responsible for the network's decision. The most important features of an input are often referred to as the discriminant region, because this is the region that the network focused on to discriminate between the PF/AR classes. If the network was trained to distinguish between spectra from active regions that did not lead to a flare, and active regions that resulted in a flare, then the heatmaps will highlight any hidden features of the spectral line that are particularly associated with preflare activity.

\subsection{Grad-CAM}
\label{GradCAMsection}
Since ConvNets are pattern recognition tools that store spatially coherent high-level representations of the input within their feature maps, much work has gone into understanding and visualizing the internal content of these special units \citep[e.g.,][]{Mahendran_2015, Dosovitskiy_2016}. It was recently demonstrated that the last layer of feature maps in a ConvNet have the ability to behave as object detectors when simply trained on binary classification tasks \citep{zhou_2014}. Further research showed that the addition of a global pooling layer (GAP) directly before the softmax classification layer, allowed researchers to simplify the network enough to project the weights of the feature maps back into the image space to extract the precise discriminant region responsible for a particular classification \citep{Zhou_2015}. Their technique called class activation mapping (CAM) had a distinct disadvantage in that it required an oversimplification of the models architecture to work, which might lead to compromises in the performance of the network. 

Gradient-weighted Class Activation Mapping (Grad-CAM) \citep{Selvaraju_2017} represents an improvement on CAM in that it requires fewer architectural constraints. It has recently been used to automatically identify important features in HMI full-disk magnetograms for flare prediction \citep{Kangwoo_2021}. Grad-CAM evaluates the importance of the patterns identified by the feature maps in the final convolutional layer, by monitoring the effect that small perturbations to these maps have on the output prediction score for a particular input spectrum. A low resolution saliency map is then constructed by taking a weighted linear sum of the feature maps, with more weight assigned to those maps that effect the output most. The saliency map is then extrapolated to the dimension of the input and projected back onto the spectrum. One weakness of Grad-CAM is that the quality of the explanation depends strongly on the selection criterion of the network’s architecture. For a detailed description of this method see appendix \ref{GradCAMsection_detailed}.

\subsection{Expected gradients}
\label{EG_section}
Expected gradients (EG) does not require special convolutional layers, although the same network used to generate explanations for Grad-CAM can be reused. Instead of perturbing components within the network to evaluate the importance of patterns, EG works directly at the resolution of the input by removing pixels/wavelength points and evaluating the net effect on the prediction score \citep{Erion_2021}. To ensure the fair distribution of pixel importance, EG makes use of a game theoretic quantity call the Shapley value $\phi_i$ \citep{Shapley1951}, which was originally constructed to fairly distribute rewards over coalitions of players. Under this formalism, the importance $\phi_{\lambda}(\mathcal{F}_\Theta)$ of a single intensity point with the current model is obtained by calculating the average difference between the prediction score under all combinations of pixel coalitions with and without the pixel in question. Since this requires taking many subsets over the input, the formalism has to be adapted to be amenable for NNs, which require information to be pumped continuously through each pixel. In-order to generate pseudo-subsets, EG "turns pixels off" by flooding them with information from the actual dataset, such that the expected value from these pixels after many passes results in a net-zero effect on the output prediction score.

Unlike Grad-CAM, the saliency maps generated from EG are often less smooth. Since it is sensible to require that the variance in importance between proximal pixels be small, EG can incorporate the attributions into the training procedure as a differentiable prior that often leads to smoother saliency maps and faster convergence \citep{Erion_2021}. We however use the attributions derived in Grad-CAM as a guide for applying a post-hoc smoothing after training. For a detailed description of EG see appendix \ref{EG_section_detailed}.

\begin{figure}[htb]
\includegraphics[width=0.5\textwidth]{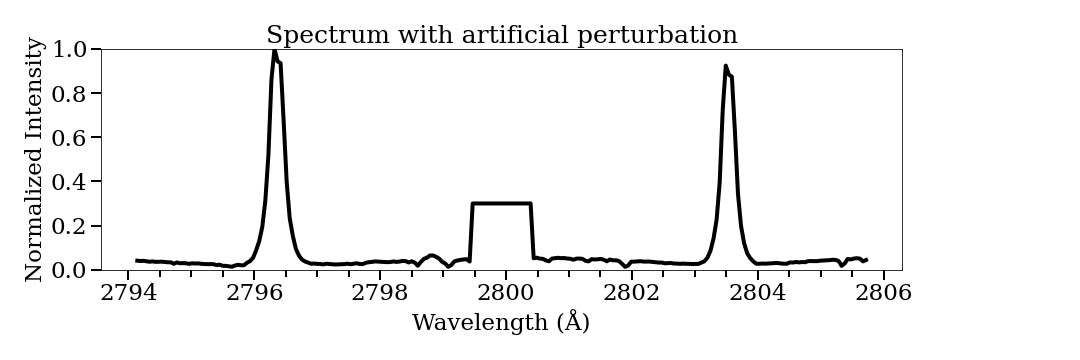}
\includegraphics[width=0.5\textwidth]{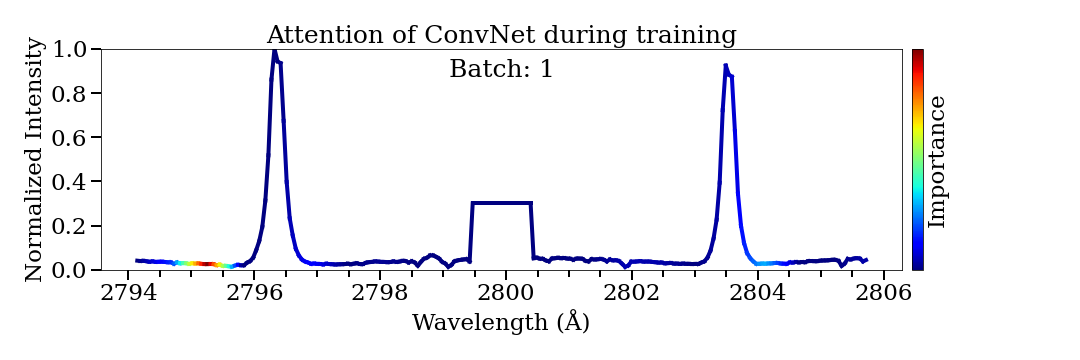}
\includegraphics[width=0.5\textwidth]{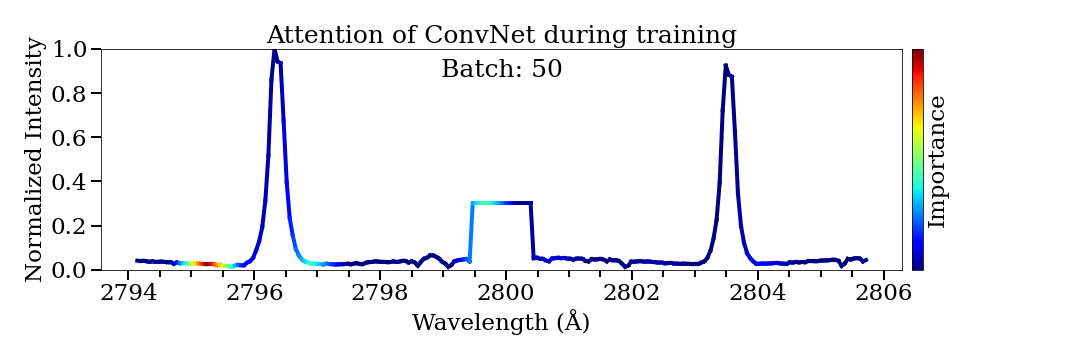}
\includegraphics[width=0.5\textwidth]{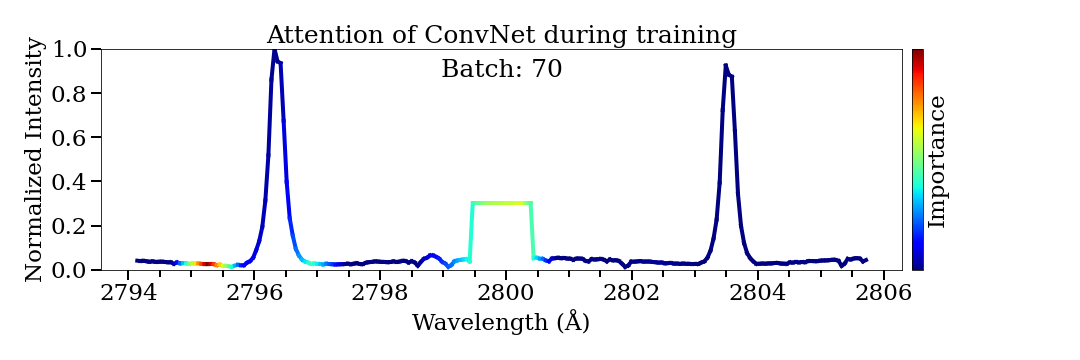}
\includegraphics[width=0.5\textwidth]{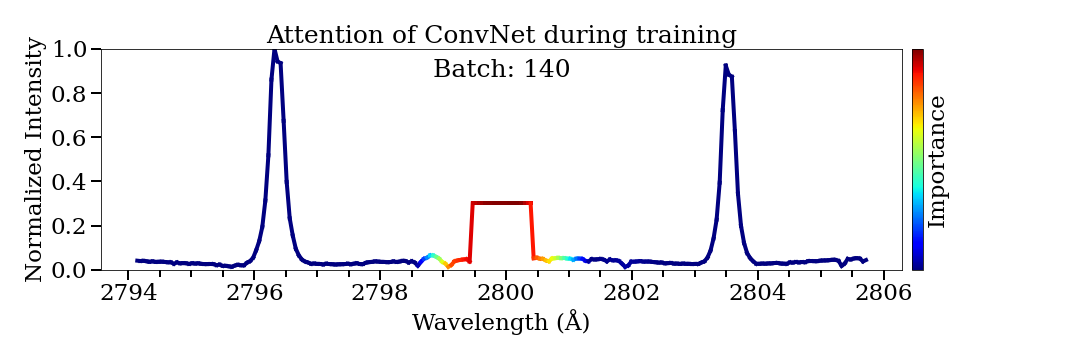}
\caption{Grad-CAM as a function of training time. A single batch means that the network has trained on $240$ spectra from the artificial dataset. Initially the ConvNet focuses randomly on the left wing of the k-line core, before pooling all of its attention onto the artifact, resulting in a classification accuracy of $100$\%. Grad-CAM allows us to deduce that the most important feature for the network's decision making was exclusively the artifact, which by design was the only way to distinguish the two classes of spectra.}
\label{evolution}
\end{figure}

\section{Testing our architecture and XAI implementations on an artificial dataset}
\label{artifact_section}
Only a few examples of attribution methods exist for one-dimensional inputs such as spectra \citep[e.g.,][]{Wang_2016}, we therefore had to test our implementation of the ConvNet's architecture and XAI methods in a controlled environment. We note here that both XAI methods returned similar results (as will be shown in section \ref{eqivv}), and as such, we only show the outputs derived from Grad-CAM.

Our artificial dataset consisted of $700$k randomly selected \ion{Mg}{II} spectra from several solar flares. From these spectra, we constructed two classes, the first class was composed out of the original flare spectra, while the second (positive) class consisted of an identical copy of the first except for a small artificial perturbation inserted at the center of each spectrum, as seen in the top panel of Fig.~\ref{evolution}. The task of the ConvNet is exactly the same as our present task, namely to solve the binary classification problem of correctly labeling the class of each spectrum. Since both classes are identical except for the perturbation, the only way for the ConvNet to distinguish between spectra from the two classes is to focus all of its attention on the perturbation.

This is exactly what we see in the remaining panels of Fig.~\ref{evolution}, which show how the ConvNet adjusts its focus during training. From top to bottom, each panel shows the heatmap derived from the Grad-CAM procedure outlined in the previous section (EG results are equivalent), with warmer colors (oranges, reds) representing wavelengths that the network has deemed important for its classifying task. The batch number associated with each of the plots indicates how many spectra the network has trained on ($1$ batch is equal to 240 spectra, and each batch represents a single adjustment to the weights). The weights of the network, both in the fully connected layers and the six filters, are initiated randomly, and as a consequence, the second panel indicates that the network has not discovered any important features. After batch $50$, gradient descent has allowed the filters to start identifying a portion of the central perturbation. With increasing training time, eventually all of the network's attention is focused on the artifact, resulting in a classification accuracy of $100$\%. This implies that 1) we have a satisfactory network architecture at least for solving simple problems, and 2) that our attribution method highlights the feature that distinguishes the two classes.\\

\begin{figure}
\includegraphics[width=.5\textwidth]{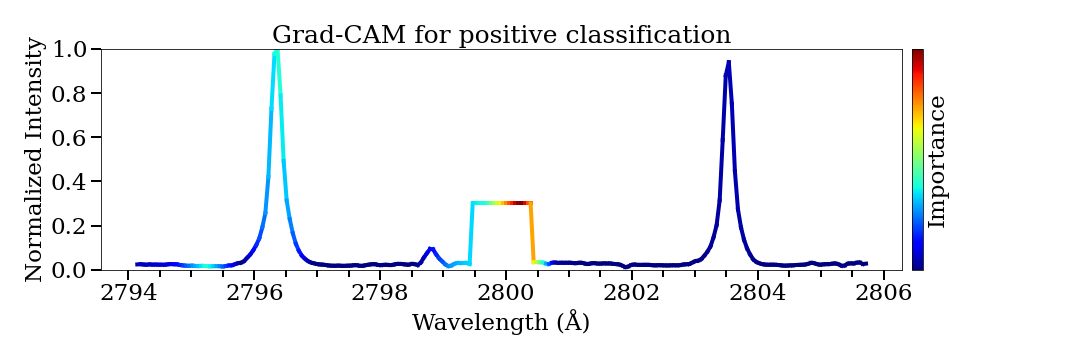}
\includegraphics[width=.5\textwidth]{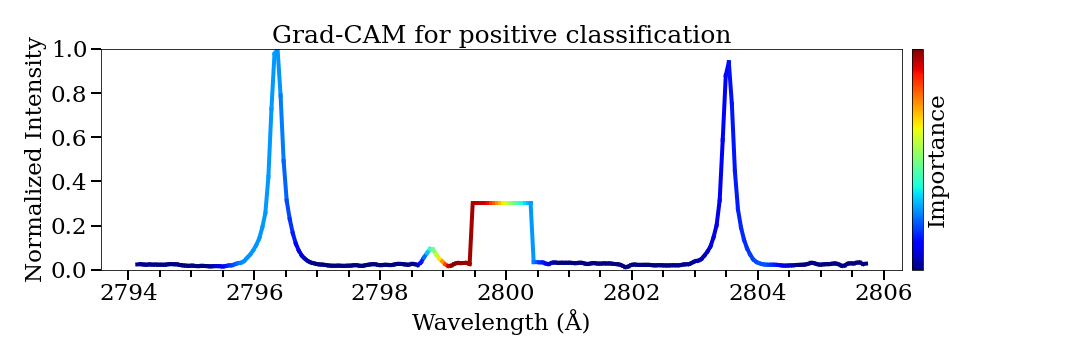}
\caption{Grad-CAM after two separate training runs. Both figures show saliency maps derived from fully trained ConvNets on the artificial dataset. In both cases, the models achieve near perfect classification scores, however, the central artifact is asymmetrically covered by the network's attention, and residual unimportant features such as the line cores still show nonzero contributions.}
\label{degen}
\end{figure}

There is however an important caveat which is not explored by this example. As it turns out, once the network has discovered a portion of the artifact its classification score saturates rapidly, and there exists very little motivation for the network to further adjust its weights to make the perturbation its central focus. This results in saliency maps (as seen in Fig.~\ref{degen}) that highlight residues of unimportant features, such as the h\&k-line cores, and partly, or asymmetrically cover the artifact. To be clear, the saliency maps shown in Fig.~\ref{degen} were derived from fully trained ConvNets, which in both cases performed perfectly on the classification task. 

\begin{figure}[htb]
\center
\includegraphics[width=0.4\textwidth]{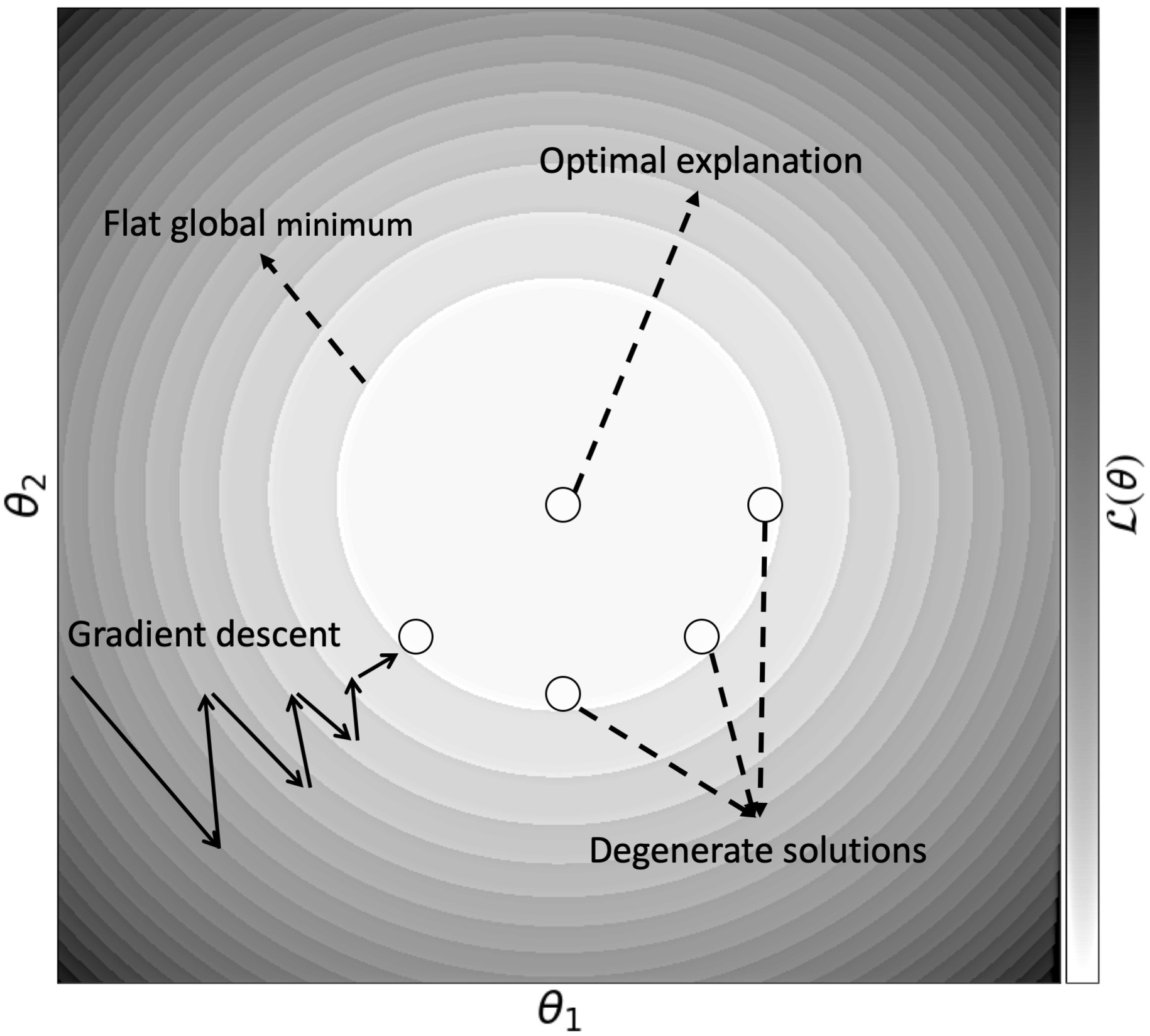} 
\caption{Schematic indicating the problem of obtaining the optimal solution both in terms of model performance and explanatory power. The x- and y-axis represent the parameters of a fictitious two parameter model. Adjusting the model parameters affects the loss, which we ideally want to minimize (lighter regions). Gradient descent allows us to converge to the global minimum in a controlled manner, however, there might be multiple solutions (as indicated by the white circles) that have the same loss. Although these points are degenerate in terms of model performance, they do not all have the same explanatory power. We postulate that the best explanation is located at the center of gravity of this degenerate space. This point must be located stochastically by initiating several models and then taking the average.}
\label{Ensemble_idea}
\end{figure}

The problem here can possibly be attributed to a flat/wide global minimum within the loss space as depicted in Fig.~\ref{Ensemble_idea}. In this hypothetical loss space, each of the white circles represent a convergent set of parameters $\{\theta_1,\theta_2\}$ found by training different randomly initiated models via gradient descent. Although each converged white circle represents an optimal solution given the loss function $\mathcal{L(\theta)}$, such as those found in Fig.~\ref{degen}, it does not represent the optimal or most intuitive explanation. In other words, when trying to derive explainable solutions we are forced to solve two optimization problems. The first is simply the loss given by Eq.~\ref{BCE}, which promotes accurate predictions, while the second loss is harder to formulate and represents the solution that delivers the maximum amount of explanatory power. Given the simple configuration of this test, we know that the best explanation is to have the network symmetrically highlight only the central artifact. This would represent the solution at the center of the well in Fig.~\ref{Ensemble_idea}.

Since the meaning of "explainable" is not easily formulated mathematically, we cannot construct an additional term in the loss function that gives the ConvNet a set of instructions to automatically transverse the degenerate space of solutions towards its center. We can however locate the center by means of a Monte Carlo approach, by taking advantage of the randomness of the initial states, and then appealing to the law of large numbers. Our course of action is therefore to generate a "swarm" of models, whose solutions populate the perimeter of the degenerate minima, and then take the average saliency map of the entire ensemble

\begin{figure}
\includegraphics[width=.5\textwidth]{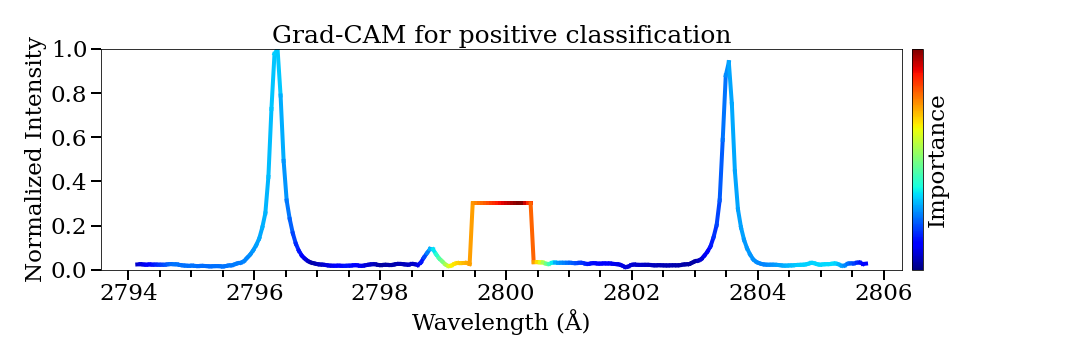}
\caption{Grad-CAM results after averaging over an ensemble of fully trained models. Although the residual attention in the cores has not fully drained to the artifact, taking the average saliency map from multiple converged runs results in a symmetric covering of the perturbation, indicating that the ensemble method offers a possible mechanism for identifying the most intuitive explanation.}
\label{Ensemble_res}
\end{figure}

\begin{equation}
\bar{M}^c = \sum_{\Theta\in\Omega} (\mathcal{F}_\Theta\to M^c)/|\Omega|,
\label{ensemble_solution}
\end{equation}
where $\mathcal{F}_\Theta$ as usual is our function parameterized by a ConvNet with unique weights $\Theta$, and $M^c$ is the attribution map produced by a particular model under the Grad-CAM formalism. The final weight solutions are part of a larger set of possible weights $\Omega$. The results of this method which we call the ensemble method, can be seen in Fig.~\ref{Ensemble_res}.

Although the line cores still occupy a residual focus of the network, the central minimum is indeed the focus of attention and symmetrically covered. The unwanted attention attributed to the cores (as seen by the turquoise coloring) can possibly be removed in several ways 1) by including more models into the ensemble, 2) by using a smart optimizer such as Stochastic Weight Averaging (SWA) \citep{Izmailov_2018} that allows the solution to stochastically wander around the central minimum before taking the average of the model weights, or 3) by implementing a sort of annealing by randomly perturbing the weights around the minimum and reducing the importance of those weights which affect the model's accuracy least. We selected the first option of including more models, although this is computationally more expensive, it has the least amount of unknowns. It should also be pointed out that in effect, the attribution prior (and less so the post-hoc smoothing) used in EG and discussed in section \ref{EG_section}, is an attempted mechanism for traversing the degenerate minima to the center of gravity.

In conclusion, the need for explainable models places a precedence on the degenerate solution space, since although all solutions have the same performance, they do not have the same explanatory power. We make the assumption that the optimal solution, both in terms of performance and explanation, is located at the center of gravity of the degenerate minimum. Due to the nonmathematical formalization of "explanation," this center of gravity cannot be searched for via controlled optimization techniques, but must be located stochastically via a Monte Carlo approach. It is not clear to what degree the problems encountered here carry over into the real AR/PF dataset, however, we expect that the loss surface will not be as degenerate, allowing the residual attention to naturally drain with increased training time. Furthermore, it is difficult to imagine why the tenants and practices derived here would not hold in more complex scenarios.

\section{Creating train and test sets}
Our data, as well as the objective of obtaining not only high performance classifiers, but also explanations from our network, complicates the training procedure outlined in section \ref{general_procedure}. The last section has clarified that we need to train multiple randomly initiated models simultaneously to derive the most intuitive set of explanations. Although we remain true to the general tenants already outlined, such as minimization of our loss function and splitting our data into a training and test set, it is not immediately obvious how one selects the data that should be in each of these sets. We also note that we do not optimize our model's hyperparameters with the intent of increasing performance metrics. This implies that a validation set is unnecessary. In this section, we explain the method and logic behind our particular choice of data partitioning.\\

Our main objective is not simply to discover that spectra from a single PF region are different from spectra from a single AR. This might be the case, but it also might simply be a peculiarity of the two selected observations. In such a circumstance, the results would not serve our purpose of understanding general flare precursory behavior, and would represent a type of overfitting. We instead wish to extract only those differences that persist over many observations. It is clear then that we require many IRIS observations, and furthermore, that we test the model on data that was not used during training. Additionally, to avoid another potential form of overfitting, spectra from the same IRIS observations cannot be both in the training and test sets, since the models could then learn particulars about the observations themselves, thus suppressing the model's ability to extract general precursory flare behavior.

To decide how to partition the 19 PF and 18 AR observations into two sets, we had to make an experimental design decision. 1) We could either randomly partition the observations, 2) perform a computationally expensive k-fold validation technique which averages over many partitions, or 3) select the partition that leads to the highest model performance. We dismissed the first option on the grounds that it could result in an unfortunate partitioning given the low number of observations, thus flooring the entire experiment. The second option, although thorough, is extremely computationally expensive and has the potential of suppressing common preflare signatures. The computational costs of this option can easily be appreciated if we consider that for every fold, one has to train multiple models to obtain optimal explanations. These models then need to generate a collection of saliency maps at a nonnegligible expense that are then averaged. These averaged maps themselves have to be averaged over all k-folds.

We therefore selected the third option based on the following logic: If a common flare triggering mechanism exists and is noticeable in the UV, then it would be presumptuous to expect that it exists in every observation, and furthermore, even if it did, IRIS's slit may sometimes miss the preflare signatures owing to its limited field of view. Therefore, to give our models the best chance of isolating a frequent preflare signature, the partitioning that leads to the highest model performance would likely coincide with a partitioning that divides observations with this mechanism into both sets.

\begin{figure}[htb]
\includegraphics[width=0.5\textwidth]{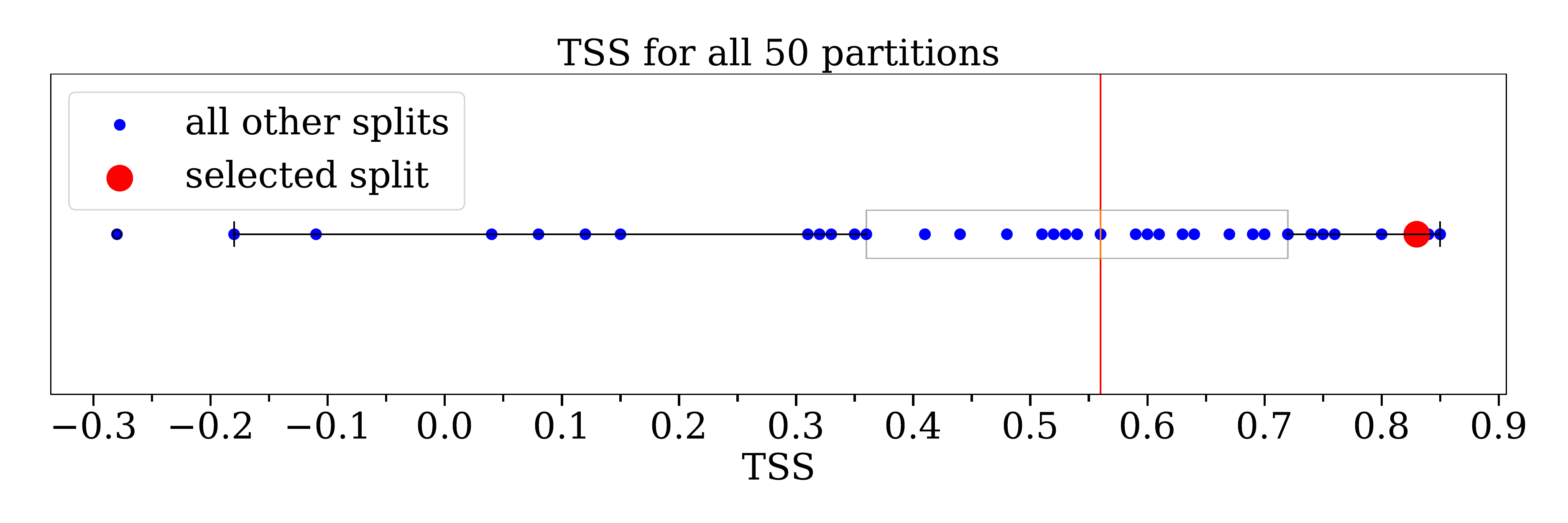}
\caption{Boxplot of the ConvNet's performance in TSS for all 50 random partitions of the observations. The red mark represents the performance of the selected model, while the blue dots are the scores from the remaining partitions. The vertical red line is the median score of all splits, while the edges of the gray box outline the first and third quartiles, the whiskers at the end denote the min and max values while anything outside of this limit is considered to be an outlier. The plot shows that the model easily achieves a high skill on the binary classification task.}
\label{ConvNet_stats}
\end{figure}

To determine this optimal partitioning, we trained the ConvNet in Fig.~\ref{ConvNet} for 50 random splittings of observations, assigning $5$ AR and $5$ PF observations to the test set and the remaining observations to the training set for each split. Since each observation contains a different number of spectra, we ensured AR/PF balance by undersampling the majority class. We note that misrepresenting the natural frequency of the classes could result in an inflated false positive rate \citep{Woodcock1976,Bloomfield2012,Deshmukh2022}, however, we found no significant effects due to undersampling. We also enforced a $3/2$ split for the final training and validation sets, meaning we had $50$\% more data for training than testing. For each of the $50$ splits, we allowed the ConvNet to train over $100$ epochs and for each case extracted the model with the lowest loss on the test set, as seen in Fig. \ref{learning_curve}. The statistics for the TSS scores of all $50$ partitions can be seen in the boxplot of Fig. \ref{ConvNet_stats}, which shows that the models in general perform skilfully on the binary classification task (a TSS $>$ $0$ is better than a random guess). We then selected the partition corresponding to the red dot, which resulted in one of the highest TSS scores.

We note that the number of spectra contributed by each observation although similar is not identical, with some observations contributing more than others. To ensure our models were not biased to a particular set of observations, we tested their performance when iteratively removing single observations from the training and test set. We noted little change to the overall TSS, which supports the generality of our results.

\section{Training a swarm of models on the partition}
The high TSS score associated with the chosen data partition in the last section increases the chance that the model learns a meaningful and common set of preflare signatures that is worth extracting using our XAI techniques. From section \ref{artifact_section}, we know that the optimal explanation is most likely obtained using Monte Carlo methods by training a large number of models in concert and then taking the average attributions. We therefore trained a 50 model ensemble on spectra from the training set, with all models achieving TSS scores around $\sim0.8$ on the test set.

\subsection{Downsampling the swarm}
Applying the full 50 model ensemble to each spectrum of the combined PF/AR dataset is computationally expensive. We therefore manually selected a subsample of 9 models that best preserved the average attributions/heatmaps of the entire swarm, resulting in drastic reductions to the computational time. To test the validity of the reduced model selection, we ran tests over several hundred randomly sampled spectra, and compared the heatmaps derived from the full $50$ model swarm and those derived from the $9$ representative models. We noted little to no loss of explainability. The perceived continuity of generated heatmaps under a smart subsampling of models can be interpreted as selecting a minimum of critical points that when averaged result in a point at the center of gravity of the loss landscape. For instance, if the loss function was that seen in Fig.~\ref{Ensemble_idea}, then one could achieve the same result by densely populating the circumference of the minimum with many models, or just simply two critical models whose parameters converge to positions opposite one another.

\begin{figure}[t] 
\begin{centering}
\includegraphics[width=.49\textwidth]{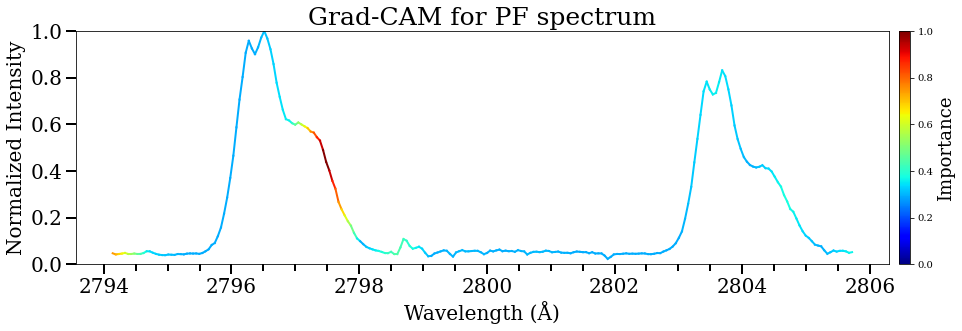} 
\includegraphics[width=.49\textwidth]{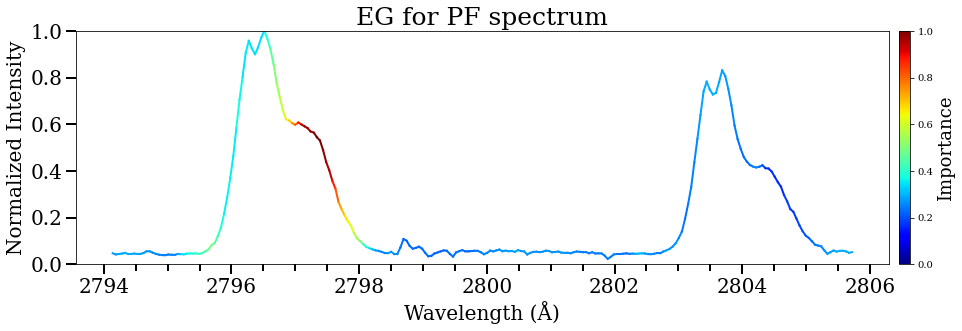} 
\caption{Attributions derived using Grad-CAM (top) and Expected Gradients (bottom). Both techniques are equivalent up to an additive constant as they highlight the spectrum in a similar way. In this case, the model indicates that red wing enhancements are important signs of impending flares.}
\label{comparison_EG_GC}
\end{centering}
\end{figure}

\begin{figure*}[tbh]
\begin{centering}
\includegraphics[width=0.49\textwidth]{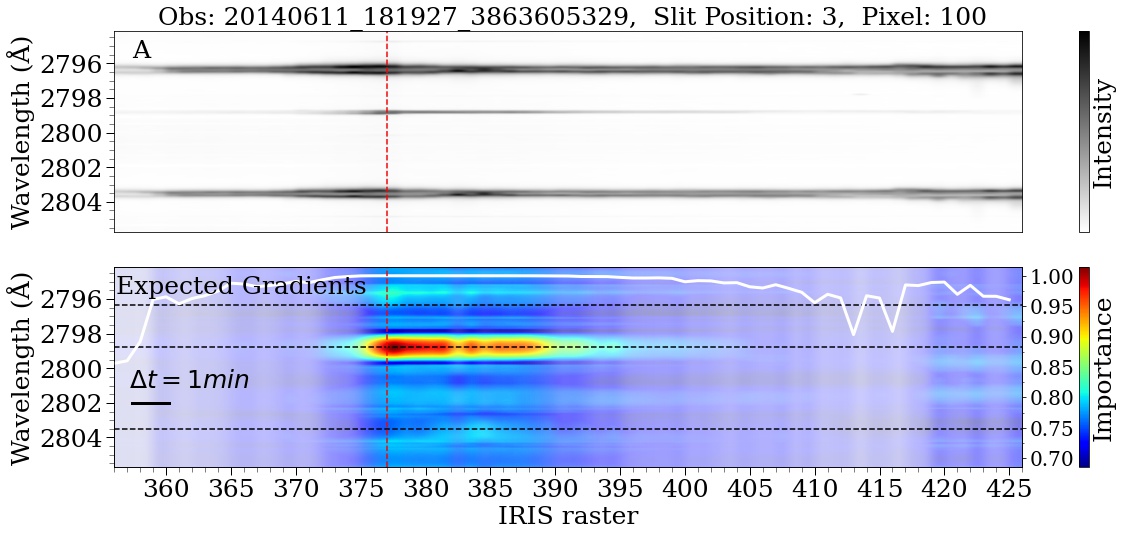}
\includegraphics[width=0.49\textwidth]{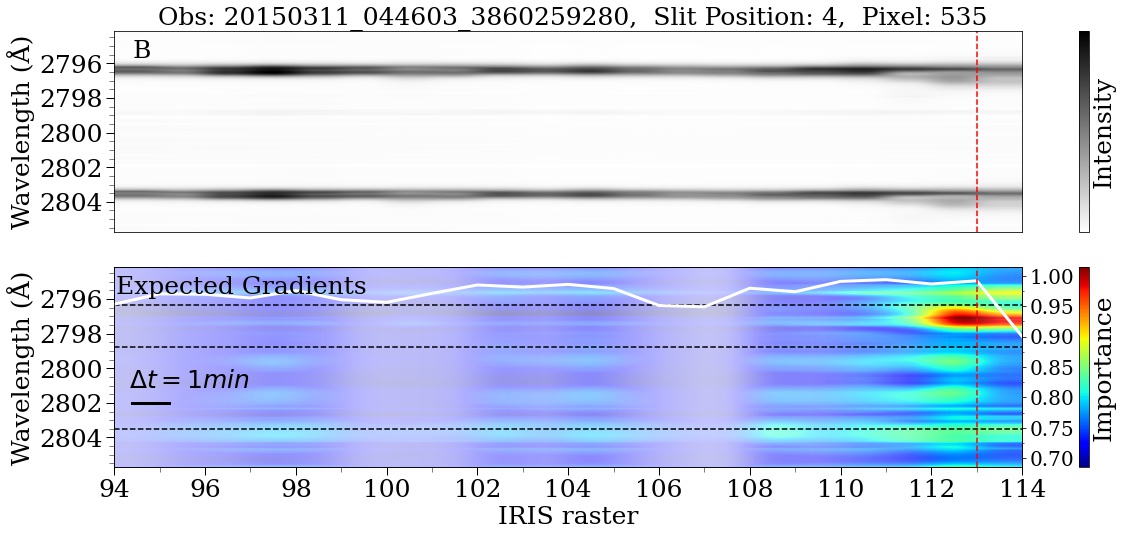}
\includegraphics[width=0.49\textwidth]{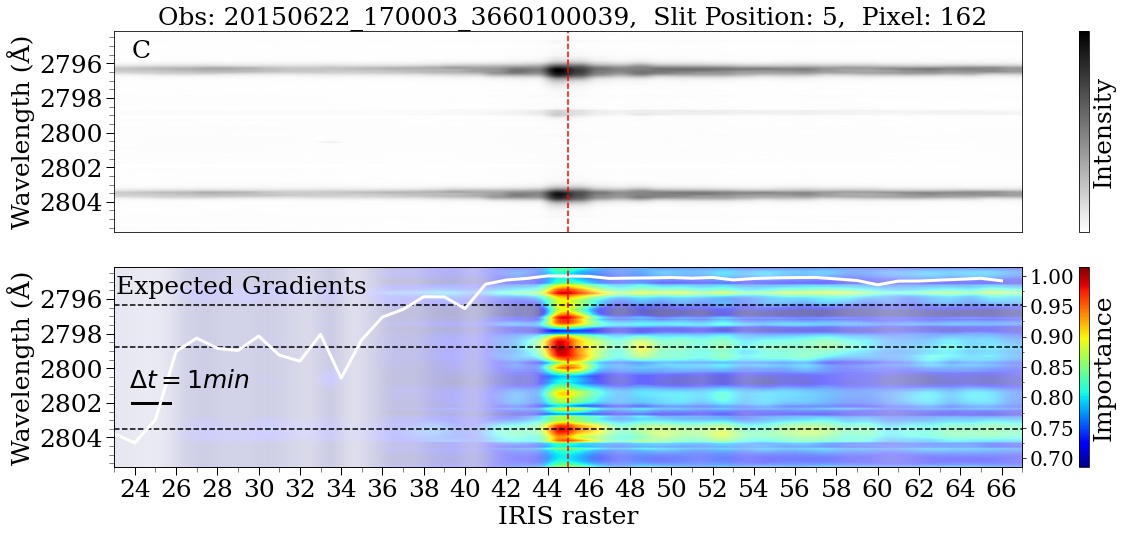}
\includegraphics[width=0.49\textwidth]{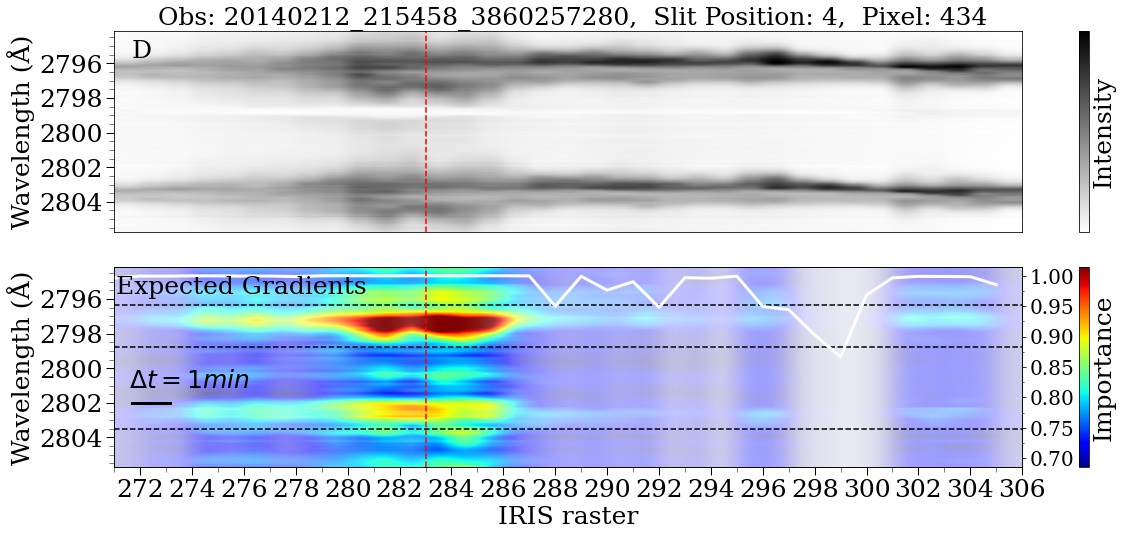}
\caption{Four spectrograms (labeled A-D) in black and white and their associated saliency maps as calculated using Expected Gradients. Darker shades in the spectrograms indicate higher intensities, while warmer colors, such as reds and oranges in the attribution maps are linked to features that the model believes are important for flare prediction. Each attribution map was normalized such that their colors are comparable. The position of the h\&k-line cores, as well as the red wing subordinate line are indicated by horizontal black dashed lines in the attribution maps. Panel A through D shows that triplet emission, downflows, broad line cores, and highly asymmetric spectra respectively, are indicative of forthcoming flares according to the model.}
\label{mosaic}
\end{centering}
\end{figure*}

\subsection{Equivalence of Grad-CAM and EG}
\label{eqivv}
We now investigate whether the Grad-CAM and EG formalisms are equivalent, in other words, whether the vector that both methods produce are similar. Figure \ref{comparison_EG_GC} shows a spectrum taken from the PF dataset and fed through the ensemble of representative models. The upper and lower panels show the derived model attributions (colors) using Grad-CAM and EG respectively. Both Explainable AI techniques highlight the spectrum in a similar way, however, the heatmap from Grad-CAM has been scaled up by an additive constant. The similarity between attributions across techniques appears consistent over different samples of spectra. We therefore conclude that the Grad-CAM and EG techniques are equivalent for our purposes up to an additive constant. In this particular case, it appears that the explanation is not optimal since the ConvNet focuses most of its attention on the k-line for its decision rather than distributing the importance evenly over both cores. This may be a consequence of the high degree of symmetry between the line core shape, such that is the network need not analyze both cores for its decision as discussed in section \ref{artifact_section}. We do however stress that these explanations are local, meaning they are particular to each spectrum, and that many times the network does treat the cores on an equal footing.

Having established this equivalence we decided to use EG for the remainder of the analysis because of its favorable axiomatic property of completeness. This property states that $\sum_\lambda\phi(x_\lambda)=\mathcal{F}_\Theta(x)$, that is, the sum of attribution values across a spectrum's wavelengths is equivalent to the prediction score of that spectrum after being passed through the ConvNet. This allows us to create an equivalence between integrated attributions and prediction scores effectively folding the one onto the other. It also means that attributions are only useful at the scale of the wavelength.

\section{Results and discussion}
In this section, we present the results of applying our ConvNet and XAI methods to the AR/PF dataset. After processing the data each spectrum becomes associated with 1) a single scalar value, which is the ConvNets prediction score $\hat{y}$ indicating whether it believes the spectrum is from the PF class (closer to $1$), or the AR class (closer to $0$), as well as 2) a vector which redistributes the total prediction score across the spectrum, functioning as our heatmap to indicate which features are most responsible for the total score. We then analyze these outputs at several progressive resolutions, firstly at the level of single spectra and spectrograms, then across the IRIS slit in time, and finally over entire observations in relation to the flare start time.

\subsection{Attributions over single spectra and spectrograms}
\label{single_spec_sec}
In order to quickly analyze what the model found important throughout the entire dataset, we compiled each spectrum into a compact representation known as a spectrogram. Spectrograms are concatenations of single spectra in time along pixels from IRIS's slit. Four examples of spectrograms can be seen in the black and white images of Fig.~\ref{mosaic}. We note that a single vertical slice from a spectrogram corresponds to a single spectrum, whose shape is encoded as an intensity map (darker here means more intense). This compactification left us with $59026$ PF and $33316$ AR spectrograms instead of millions of single spectra. We then ordered the spectrograms according to their mean prediction scores and manually scanning through the results with the highest scores, that is, those spectrograms that were strongly associated with flare precursory activity. Within these spectrograms we found consistent patterns that are well represented by the four examples shown in Fig.~\ref{mosaic}. 

Each of the four panels labeled A-D consist of the original spectrogram (with darker colors indicating more intense emission) and its associated saliency map. The saliency map consists of stacked single spectrum attributions like those in Fig.~\ref{comparison_EG_GC}. The y-axis (left) is always wavelength (measured in Angstroms) while the x-axis is the raster number of the original IRIS observation, which implicitly encodes time as a function of the raster's cadence. For clarity, a minute interval is demarcated by a black horizontal line in the attribution maps. Since all of these observations span $25$ minutes and terminate with a solar flare, the lengths of the markers are equivalent. The y-axis (right) tracks the prediction score which is traced out by white lines in the saliency maps. The position of the vacuum \ion{Mg}{II} h\&k-line cores as well as the red wing triplet emission are demarcated by black dashed horizontal lines.

The scores and attributions are encoded into the attribution maps as follows: The average score assigned to each spectrum (at each raster) from the ensemble of representative models encodes the transparency of the map, such that darker regions are seen to be more important for flare prediction. EG then takes these raw scores and fairly distributes them across the wavelengths turning scores into attributions. Regions of high attributions are colored with warmer colors and indicate those features which are seen as important for flare forecasting. All panels were normalized with an arbitrary value to make their respective colors comparable.

\begin{figure}[htb]
\begin{centering}
\includegraphics[width=0.49\textwidth]{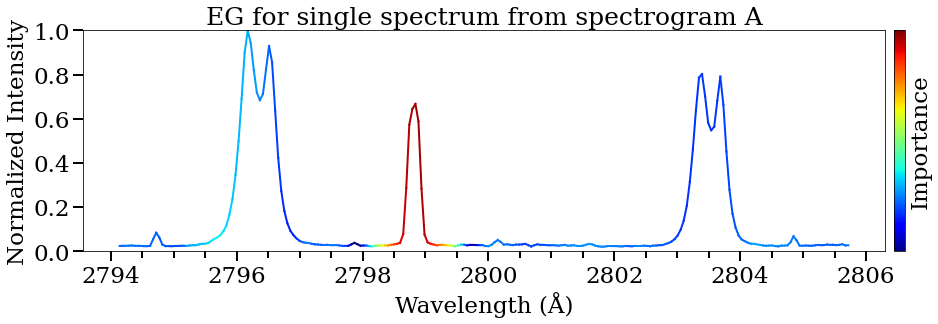} 
\includegraphics[width=0.49\textwidth]{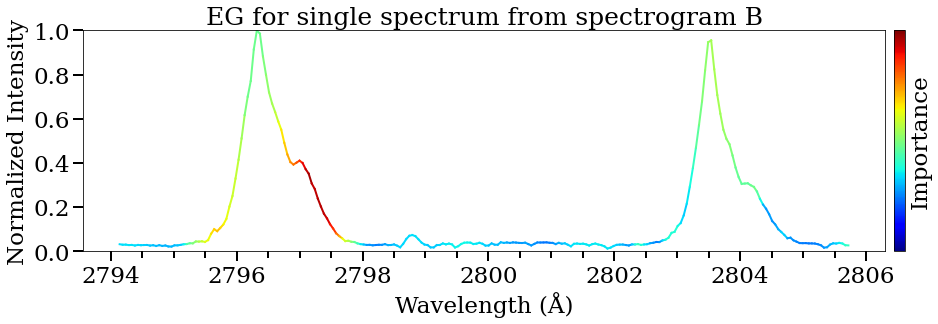} 
\includegraphics[width=0.49\textwidth]{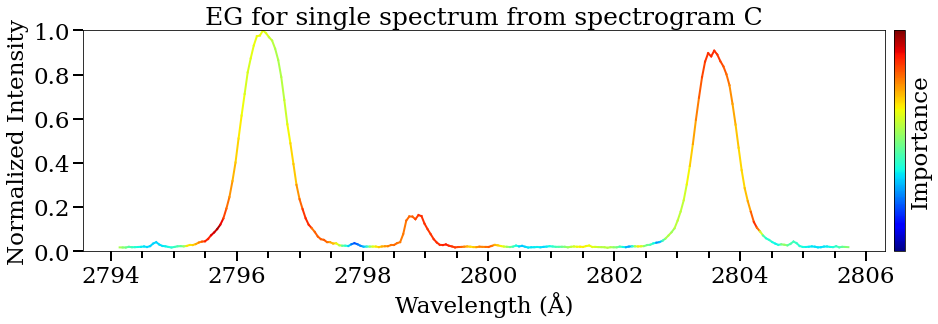} 
\includegraphics[width=0.49\textwidth]{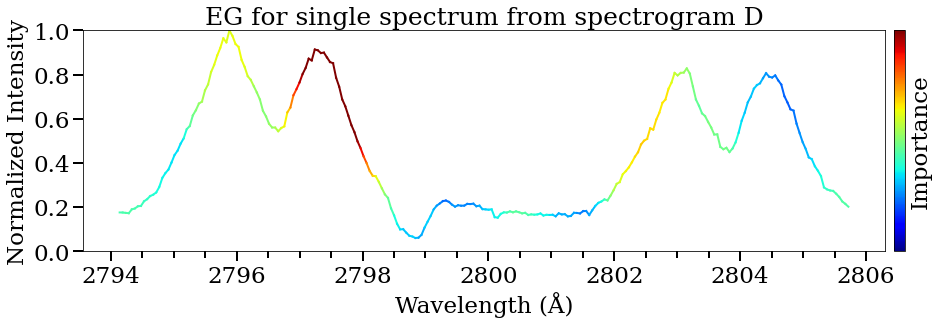} 
\caption{Attributions of individual spectra associated with the saliency maps in Fig.~\ref{mosaic}. Each spectrum is a slice from the attribution maps from the locations indicated by the vertical dashed red lines. The y-axis is normalized intensity and the x-axis is wavelength in angstroms ($\text{\AA}$). Red, orange and turquoise colors indicate features that the model associates with precursory flare activity.}
\label{mosaic_specs}
\end{centering} 
\end{figure}

Panel A shows that triplet emission is strongly indicative of preflare activity, since EG highlights the portion of the spectrogram (in red) where this emission can be seen to become enhanced. When the triplet emission subsides, the model's attention synchronously dissipates (colors turn from red to blue). Similarly, panel B indicates that velocity flows in the form of red wing enhancements are also indicative of preflare activity as can be seen starting around raster $112$. It is important to note that upflows are regularly flagged as important but are not shown in the figure. The spectra in Panel C appear to become substantially broader around raster $44$, and are consequently highlighted orange and red, indicating a positive relationship between preflare features and broad spectra. Finally, panel D shows extremely broad asymmetric spectra that are flagged as strong predictors. An example spectrum from each panel can be seen in Fig.~\ref{mosaic_specs}, each of which were extracted from their corresponding spectrogram from the locations indicated by the red vertical dashed lines. 

We note that the prediction scores, as traced out by the white lines in the attribution maps of Fig.~\ref{mosaic}, are saturated close to the maximum PF prediction score of one. This is because these maps represent some of the highest scoring maps in our dataset, while other maps scored much lower.

\subsection{Attributions over IRIS slits}
The location of high attributions (warm colors) within the 25 minute windows of each saliency map appear to be randomly distributed, and do not display a monotonic increase in prediction score closer to flare onset as was found for the case of the  X1.6-class flare observed by IRIS on 2014 September 10 \citep{Panos2020}.  It is important to note that for the September 10 event, the authors restricted the pixels of the sit-and-stare to those associated with enhanced activity within the SJIs. To conclude whether or not there is any general tendency for prediction scores to increase during flares, we have to integrate over the entire IRIS slit.

\begin{figure}[htb]
\begin{centering}
\includegraphics[width=0.5\textwidth]{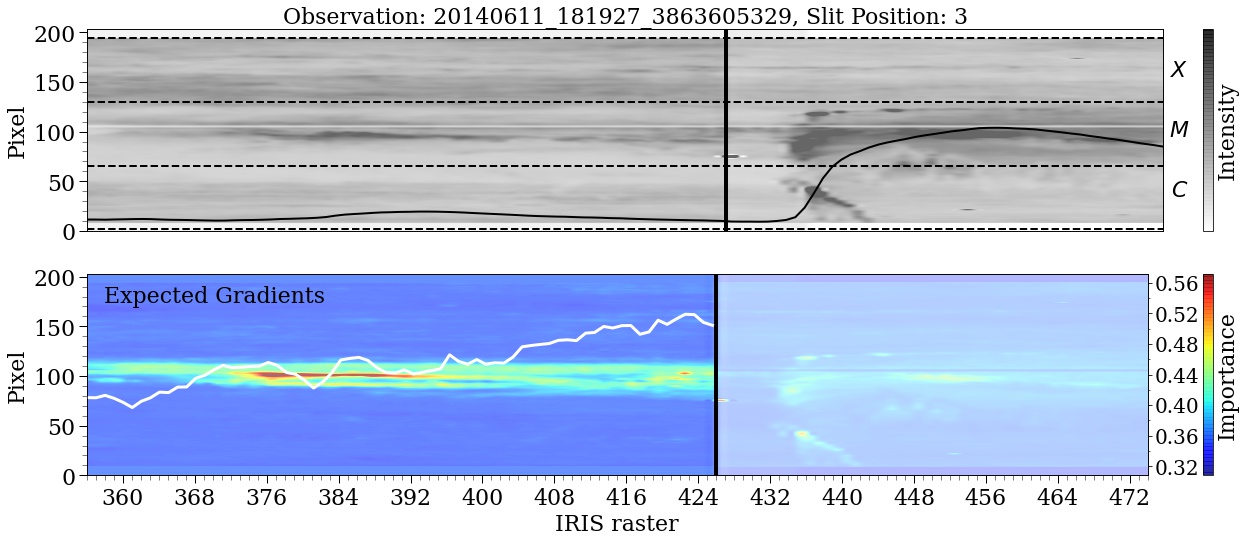}
\includegraphics[width=0.5\textwidth]{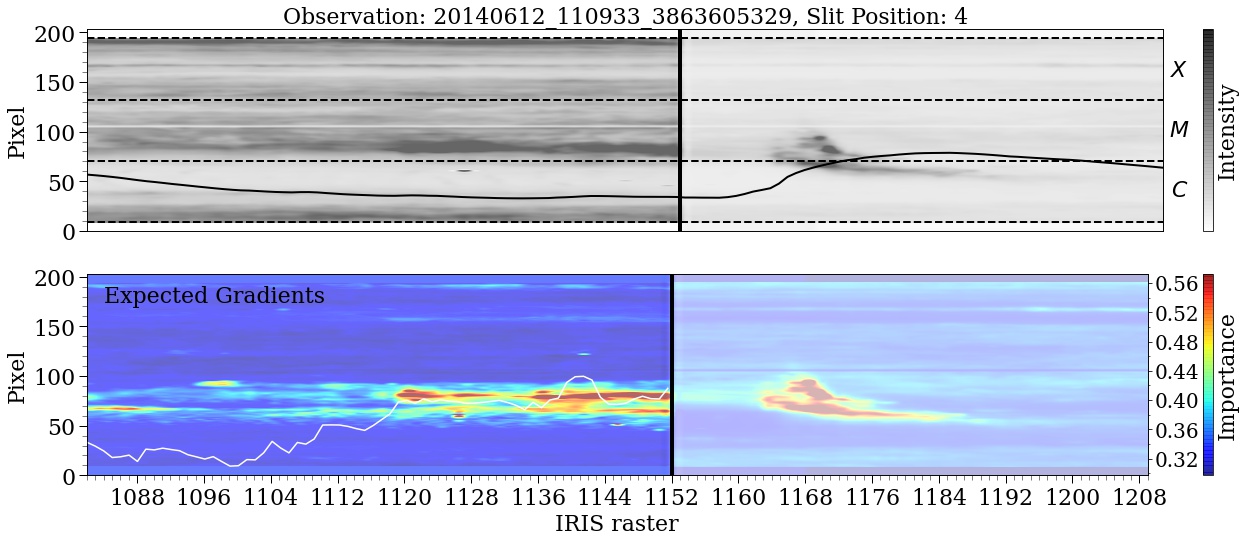}
\caption{Attributions over IRIS slits. The top panel shows an intensity map (black and white) over the pixels of one of IRIS's slit positions in time, with darker colors indicating more intense emission. The corresponding saliency map can be seen below, with warmer colors (orange, red etc.) indicating regions that are seen by the models to be important for flare prediction. The black curve in the intensity map corresponds to the GOES-curve whose flare level is indicated on the right axis, while the white curve in the attribution map is the mean prediction score as a function of time along the slit (score seen on the right axis). The bottom panel is the same but for a different observation. Both examples have an increasing prediction score closer to flare onset.}
\label{Integration1}
\end{centering}
\end{figure}

Figure \ref{Integration1} shows two positive instances that demonstrate approximate monotonic increases in prediction score with time (for a comparison with nonflare events see Fig.~\ref{Integration_neg}). Here we have taken the mean attribution score of each spectrum, that is, we integrated over wavelengths to produce saliency maps over the entire IRIS slit. The black and white panels indicate the intensity along the slit at different times with darker colors being more intense than lighter colors. The black curves show the GOES X-ray flux with flare class indicated on the y-axis on the right. Below each of these images are their associated saliency map which as usual indicates the most important regions along the slit in time using the same color code (warmer colors are more important for prediction). The white curves indicate the prediction score in time integrated over all pixels, not just a subset that was used in \citep{Panos2020}.
The upper left panel of Fig.~\ref{mosaic} is in-fact a spectrogram taken from a single pixel of the upper panel in Fig.~\ref{Integration1}, while the lower panel is of a new PF observation entirely. The vertical black line indicates the end of the PF period and the start of the flare according to the GOES catalog. The intensity and attribution maps were normalized separately across both sides of this divide, to ensure that the details of the PF period were not drowned out by the relatively much higher intensities and attributions during the flare. We find that in roughly 105 out of 135 slits from the 19 PF observations, the location of the maximum attribution along the slit is predictive and aligned with the maximum UV emission later during the flare. Another good example of this can also be seen in Fig.~\ref{aligned_pic} in the supplementary material.

In both cases shown in Fig.~\ref{Integration1}, we see that there is a general tendency for high prediction scores to be associated with enhanced intensities, even though the intensities are only implicitly encoded into our models after the normalization. If an energetic event occurs within the chromosphere, the excess emission from the line cores typically vastly outcompete the gains in the continuum, leading to a pseudo continuum that appears flat after normalizing the spectra by their maximum values, see section \ref{data_section} for details.

\begin{figure}[t]
\begin{centering}
\includegraphics[width=0.47\textwidth]{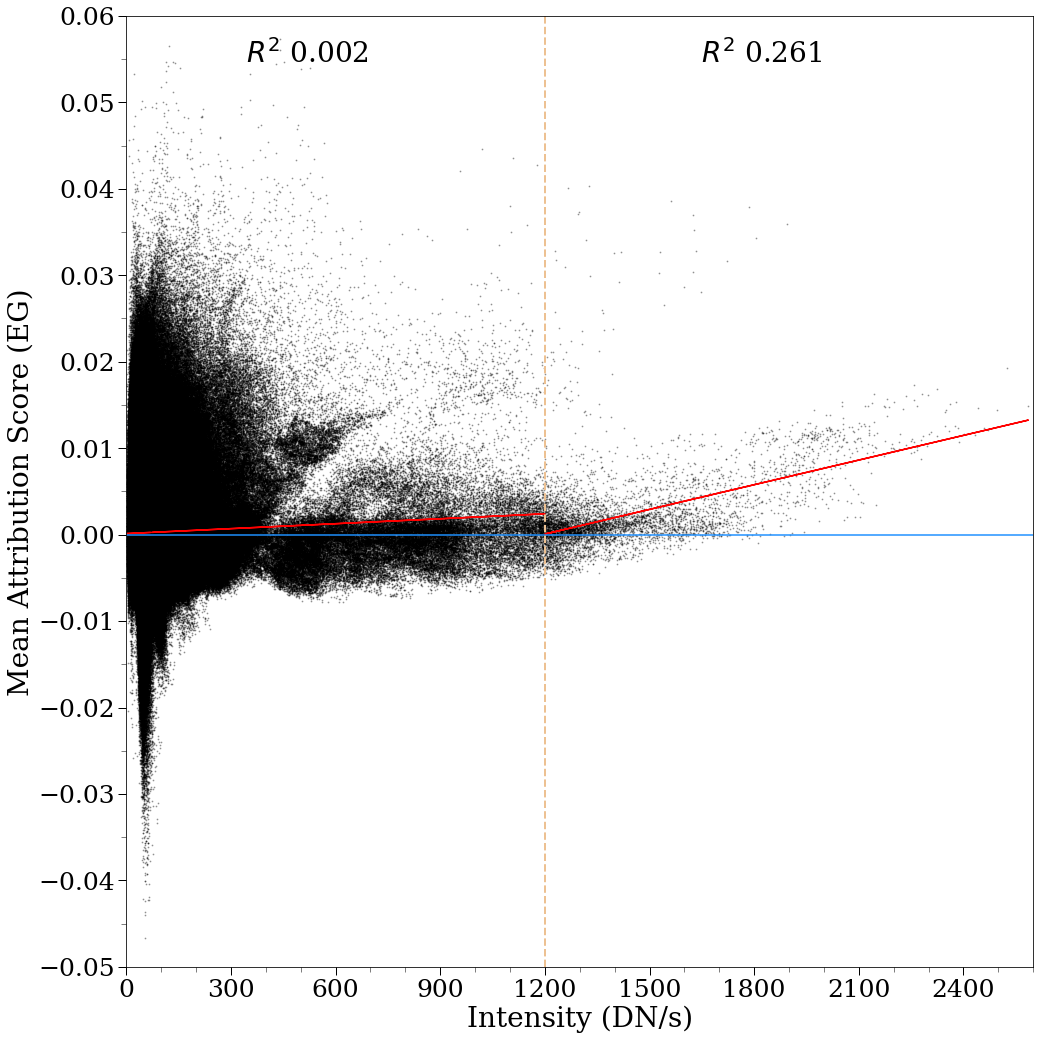}
\caption{Scatterplot of the average attribution using (EG) for each spectra in the PF dataset as a function of intensity (DN/s). The plot is divided into a low and high intensity regime by a dashed vertical line. The red lines in either regime represents the best linear fits, and the corresponding $R^2$ values are provided in the upper part of each regime. We see that "low intensity" spectra can still be critical for flare prediction, and that very high intensity spectra are more strongly correlated with higher prediction scores.}
\label{intensity_vs_attrib}
\end{centering}
\end{figure}

To examine the relationship between intensity and prediction score in more detail we plotted the mean attribution score of every PF spectrum as a function of DN's per second in Fig.~\ref{intensity_vs_attrib}. Larger average attributions imply spectra that the model believes is more useful for flare prediction. In the high intensity regime right of the dashed orange line, the intensities appear to be positively correlated with the attributions, with higher intensity spectra being associated with higher attributions, however, towards the "low intensity" regime, the coupling between attribution score and intensity is only weak to nonexistent, and it is possible to have low intensity spectra that are critical for flare prediction.

\subsection{Global explanations}
\begin{figure}[t]
\begin{centering}
\includegraphics[width=0.481\textwidth]{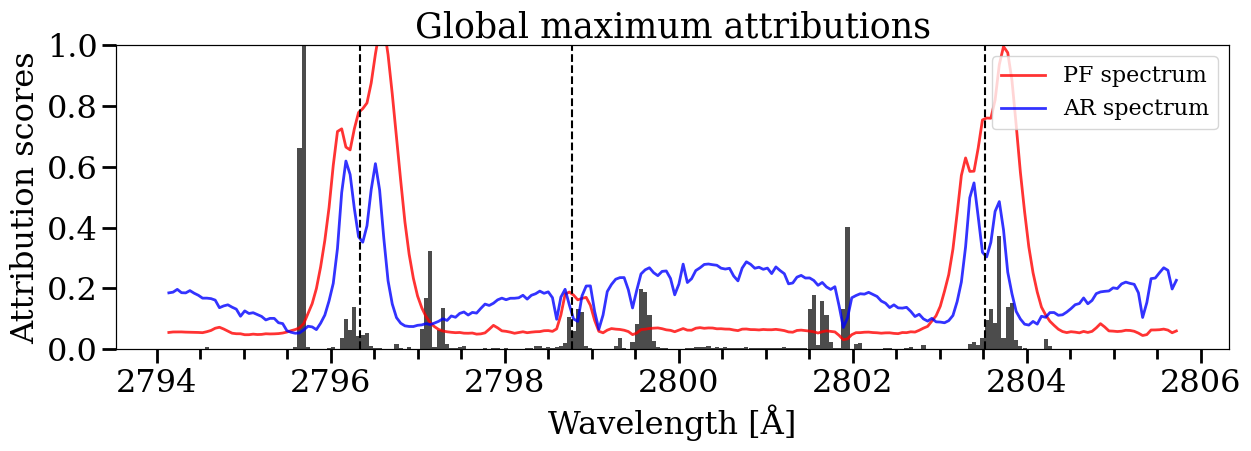} 
\includegraphics[width=0.5\textwidth]{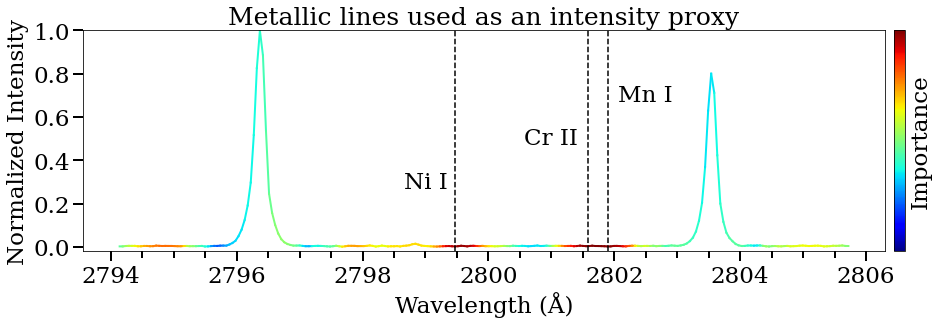} 
\caption{Global explanations. Upper panel: Distribution (black bars) showing where the model focuses most of its attention on aggregate for the spectra that have high prediction scores. For context, a red and blue spectrum from the PF and AR class have been plotted respectively, with vertical black dashed lines indicating the positions of the line cores and triplet. Flows that show up left and right of the k-core are important as well as emission in the triplet line. Bottom panel: The model focuses on the metallic absorption lines as a proxy for intensity, since intense spectra after being normalized lead to flat pseudo continuum.}
\label{global_exp}
\end{centering} 
\end{figure}

Because the explanations derived from Grad-CAM and EG are local, that is, specific for each spectrum, we aim to derive global explanations so that we can investigate PF features in general. To do so, we evaluated the distribution of attributions for the highest scoring PF spectra, that is, those spectra that had prediction scores between $0.9$ and $1$. We then determined the position and magnitude of each spectrum's maximum attribution score and summed the results, allowing us to track the aggregate importance of each wavelength for the model's decisions. The resulting distribution can be seen in the top panel of Fig~\ref{global_exp} together with an arbitrarily scaled PF spectrum (red) and active region spectrum (blue) for reference. The vertical dashed lines mark the position of the \ion{Mg}{ii} h\&k line cores as well as the \ion{Mg}{ii}  triplet emission. The histogram makes it clear that in general, the model focuses on regions to the left and right of the k-core, with some attention given to the cores themselves, the triplet emission, as well as three metallic lines. This offers statistical support for the observational claims made in section \ref{single_spec_sec} in relation to Fig~\ref{mosaic}, indicating that features such as triplet emission, flows in the form of red and blue wing enhancements, as well as broad cores all contribute positively to the network's prediction score. The large focus given to wavelengths on the left of the k-core is not exclusively due to upflows, with the model also using these wavelengths to identify broad spectral cores such as those indicated in the third panel of Fig~\ref{mosaic_specs}. In fact, upon visual inspection, downflows are far more prominently given attention than upflows in our dataset.

The model appears to also place focus on particular absorption lines in the pseudo continuum corresponding to \ion{Ni}{i}, \ion{Cr}{ii}, and \ion{Mn}{i}. To explain the model's focus on these metallic lines, we randomly plotted many samples of spectra that coincided with high attributions around these wavelengths. We found no evidence of emission in these lines, however, we know that the continuum height acts as a proxy for intensity (high intensity leads to flat continuum after normalization) and therefore we assume that the model uses these absorption lines to determine the "flatness" of the continuum and thereby the intensity of the spectra. This is supported by a PF observation where the highest attribution scores of the models are clustered around the metallic lines. An example spectrum and its associated attributions is shown in the lower panel Fig~\ref{global_exp}.\\

Interestingly, the second panel of Fig.~\ref{Integration1} shows large increases in attributions from pixels $50-100$ towards the end of the PF period where the GOES curve is relatively flat. Something to note is the decrease in prediction score, which went from maximum values of $1$ in Fig.~\ref{mosaic}, to values around $0.4$ in Fig.~\ref{Integration1}. This decrease in prediction score is a direct consequence of integrating over the entire slit, where many of the slit's pixels are sampling inactive regions of the Sun. The dependence of the score on arbitrary integrations shows that under this current framework, we have to reevaluate our metric and method of predicting flares in real-time.

\begin{figure*}[t]
\begin{centering}
\includegraphics[width=\textwidth]{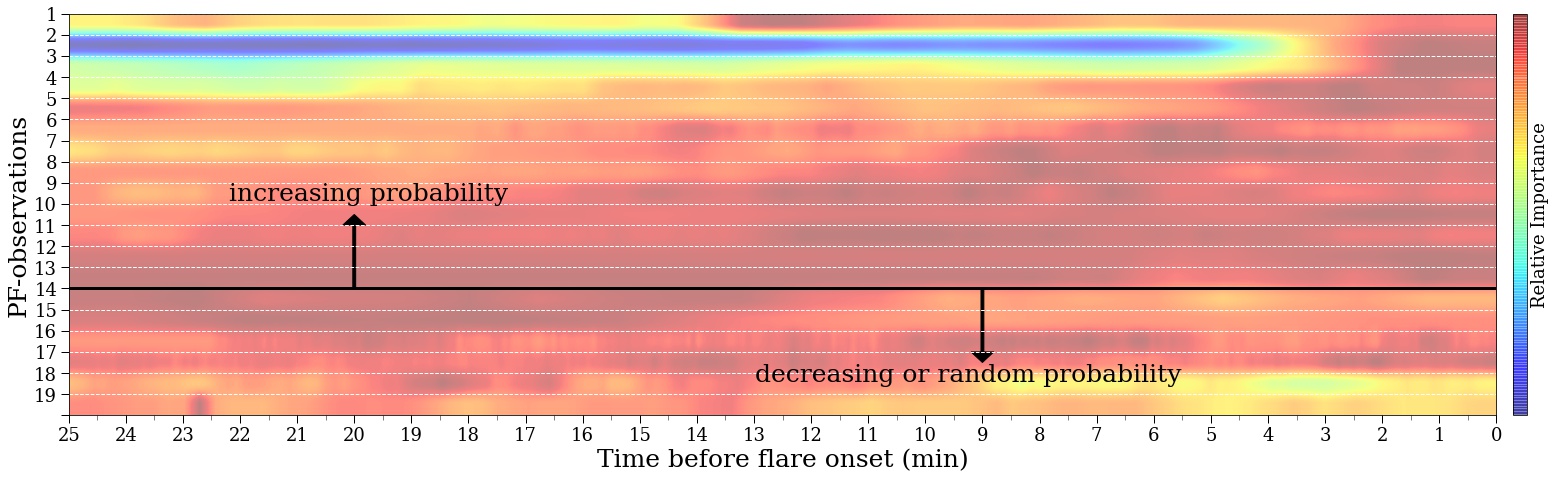}
\includegraphics[width=\textwidth]{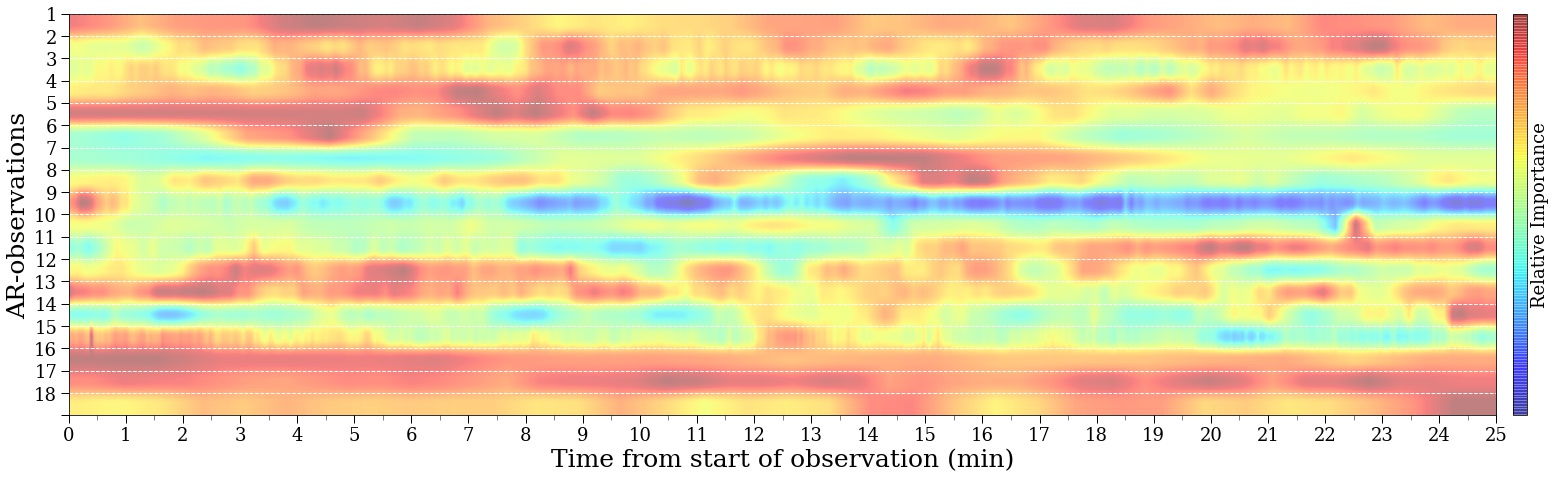}
\caption{Attributions for all observations. The top and bottom panel indicate the variation across all pixels and rasters of the models prediction score with time for all PF and AR observations respectively. Each row is normalized by its maximum value, and corresponds to a single observation. Higher relative prediction scores are indicated by warmer colors. We see that the PF observations (top) are much more structured and tend to predict flares more strongly closer to flare onset, while the AR scores (bottom), result in a heatmap that is far more sporadic in time. The numbers on each panels y-axis correspond to the observation numbers listed in Table \ref{obs_table}.}
\label{Integation2}
\end{centering}
\end{figure*}

\subsection{Monotonicity}
The problem of flare prediction is best framed as a two part problem. The first part is relatively insensitive to the time domain, and is only concerned with the performance of a model on a binary classification task. This is the approach that the majority of flare literature addresses, and it is not principally concerned with how these data are distributed in time or at what point within the time window the flare will occur. The second part concerns itself with how the prediction score evolves over time. The identification of monotonic trends in features and or monotonic increases in predictions are key to deploying practical flare forecasting strategies. 

One can imagine a scenario where triplet emission is critical for distinguishing two classes (AR/PF) because this emission is high for all PF observations and low for all AR observations, however, it might be the case that the triplet emission, although a greater indicator of class, does not significantly vary within the PF observations themselves, so that it has a very weak predictive utility along the time dimension. In summary, we need to identify features that not only tell us if a flare will occur, but also when it will occur. There is also no a priori reason to believe that these two sets of features need be the same.

\subsection{Attributions over entire IRIS rasters}
Following the above discussion, it is also possible for the IRIS slit at a particular step during the raster to be positioned off of the active region. When this happens, the prediction score does not monotonically increase, but can randomly fluctuate and even decrease with time. By calculating scores in intervals averaged over each raster of IRIS, we can set a fair baseline prediction for each observation. Since this average score is arbitrarily predicated on the active region itself as well as IRIS' coverage of the region, the baseline of one observation is not comparable to the baseline of another observation, however the trends of the prediction curve in time as well as its inclination do become comparable. 

Figure \ref{Integation2} condenses the behavior of our entire PF (top panel) and AR (bottom panel) dataset within a single figure. The upper panel shows a heatmap, where the y-axis indicates PF observation, whose number corresponds to Table~\ref{obs_table}, and the x-axis is time in minutes before flare onset. Again the warmer colors indicate periods of time that are seen by the models as important for predicting flares. In other words, redder colors imply stronger warning signals that a flare will occur. 

To derive these heatmaps, each flare observation was placed on a uniform time grid with intervals of 1 second and spanning the PF period of $25$ minutes. A matrix of size $(19 \times 1500)$, corresponding to $19$ PF observations and $1500$ seconds, was then incrementally populated with the average prediction score over an entire raster for each observation. Since most observations typically have cadences of more than two seconds, the initial high cadence grid (and corresponding matrix) was only sparsely populated. Missing values were then filled in by linearly interpolating across the calculated scores within each observation. Each of the $19$ rows of the matrix were then normalized by their maximum value so that observations with relatively high baseline prediction scores did not drown out the behavior of prediction curves from observations with lower baseline scores. Furthermore, the observations were purposefully ordered both within Table \ref{obs_table} and the matrix, such that observations with similar normalized prediction curves would appear close together, thus aiding the identification of patterns across the dataset. The exact same steps were also carried out for the AR observations seen in the bottom panel.

We note that the PF dataset shows much more structure than its AR counterpart which appears to have normalized prediction curves that randomly evolve with time. The PF observations as seen in the upper panel, can be divided into two regimes (a strong and weakly predictive regime) depending on the behavior of their prediction curves. Above the black horizontal line, we see that 13 of the 19 observations tend to have prediction curves that increase as we approach each observations flare onset. Furthermore, the prediction envelope with time smooths out as we move down from PF observation $1$ to $13$. Below the horizontal black line we have six PF observations that do not appear to display any coherent behavior with time.

\begin{figure}[t]
\begin{centering}
\includegraphics[trim=.6cm 1.3cm 2cm 2cm, width=.233\textwidth, clip]{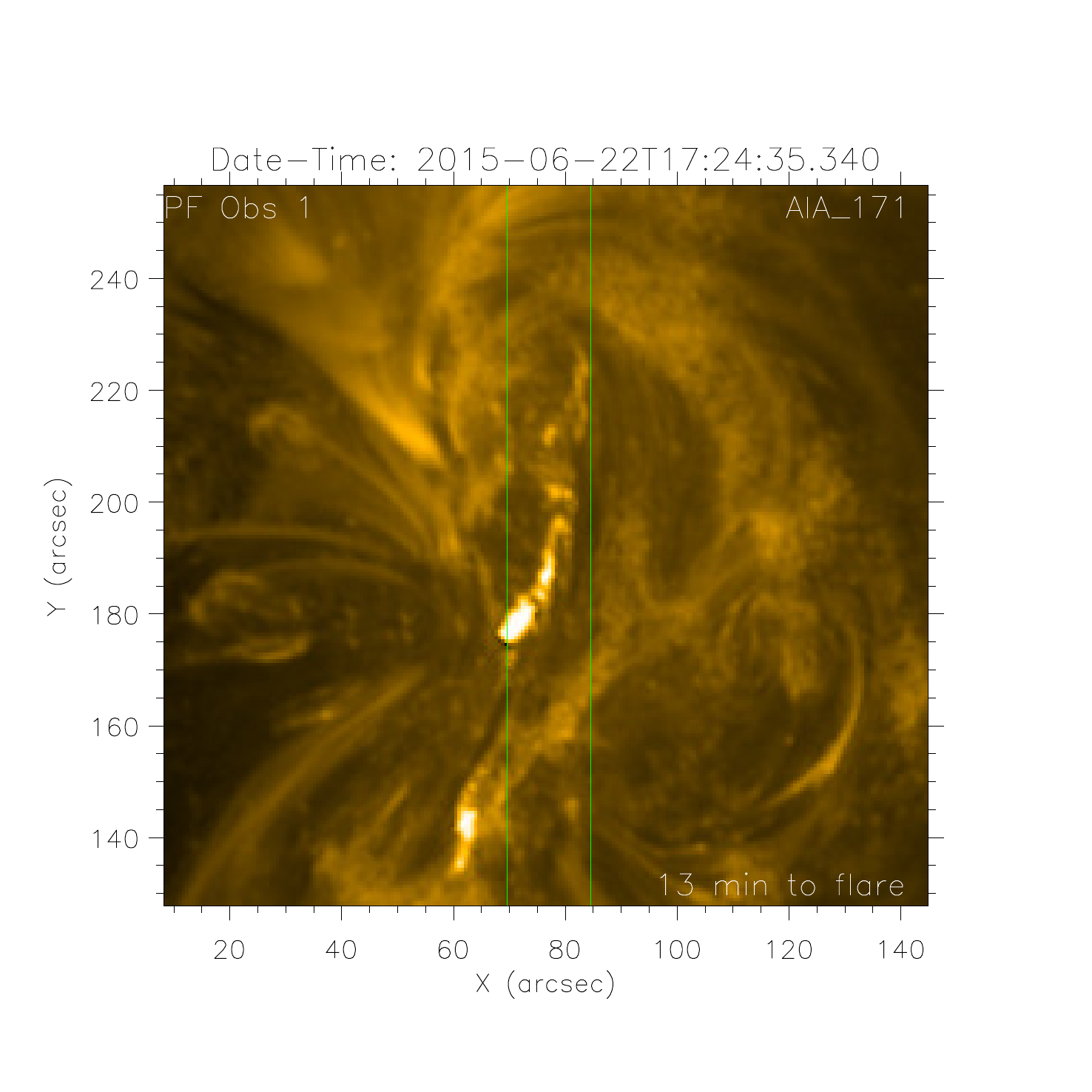}
\includegraphics[trim=.6cm 1.5cm 2cm 2.4cm, width=.233\textwidth, clip]{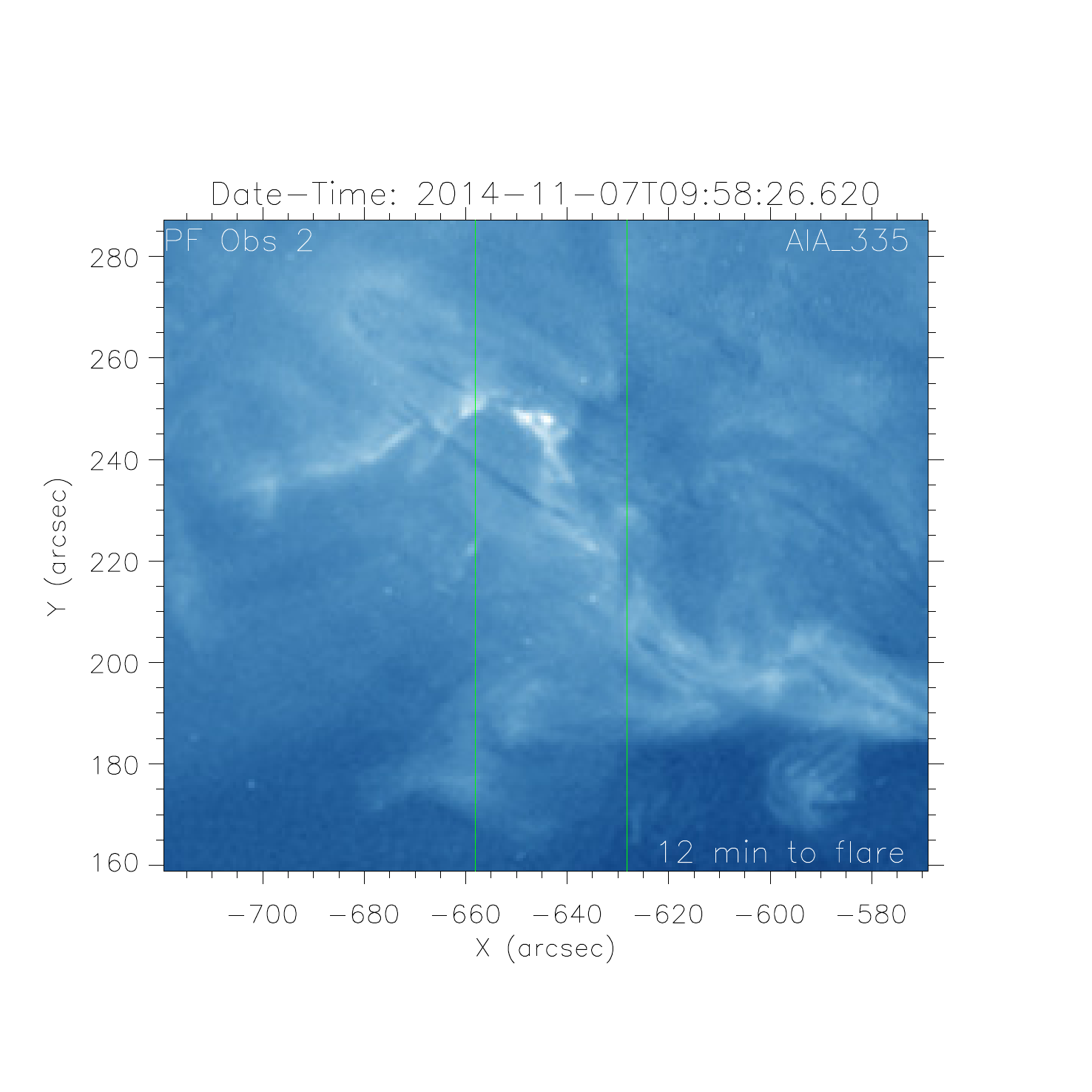}
\includegraphics[trim=.6cm 1cm 2cm 1.5cm, width=.233\textwidth, clip]{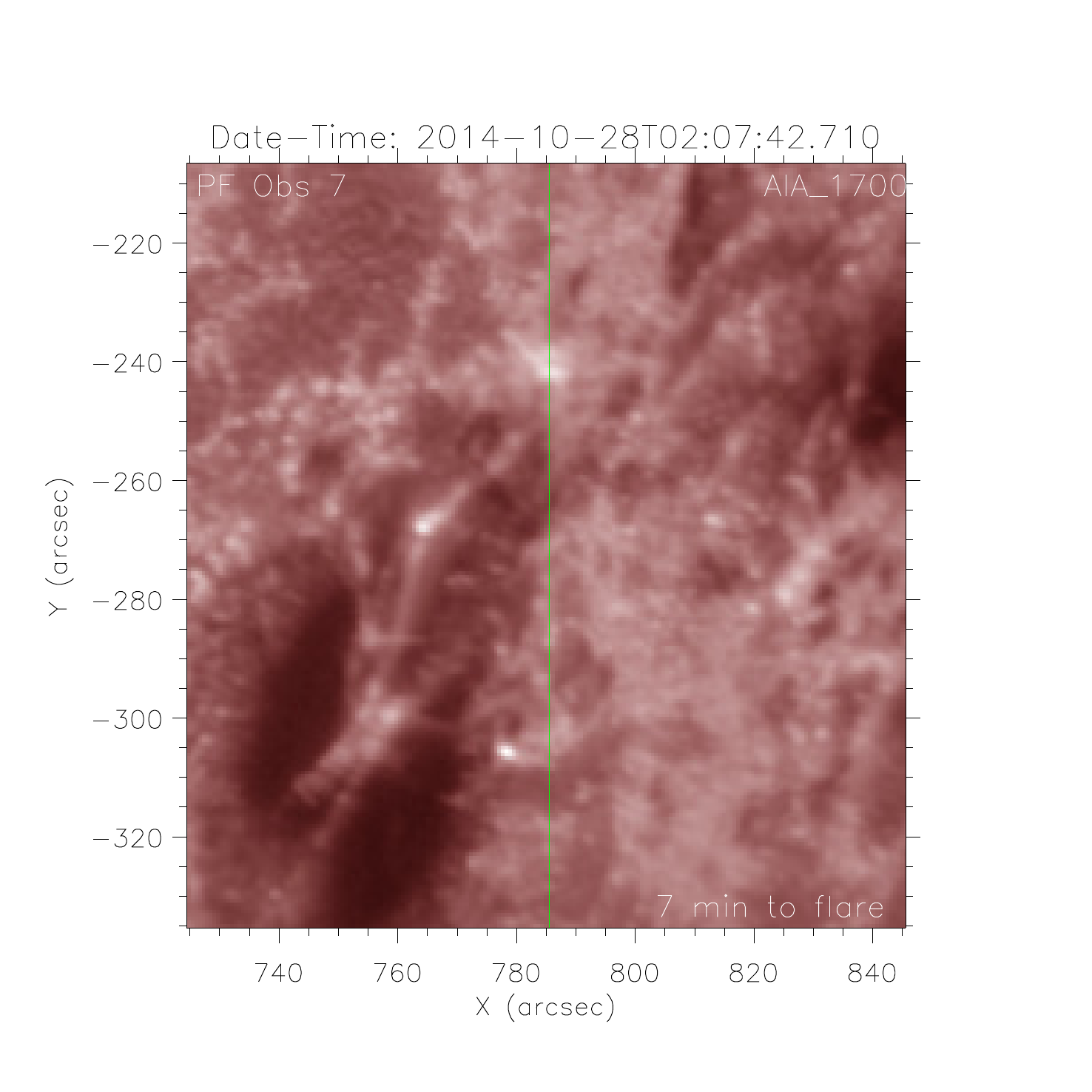}
\includegraphics[trim=.2cm .8cm 2cm 1.6cm, width=.233\textwidth, clip]{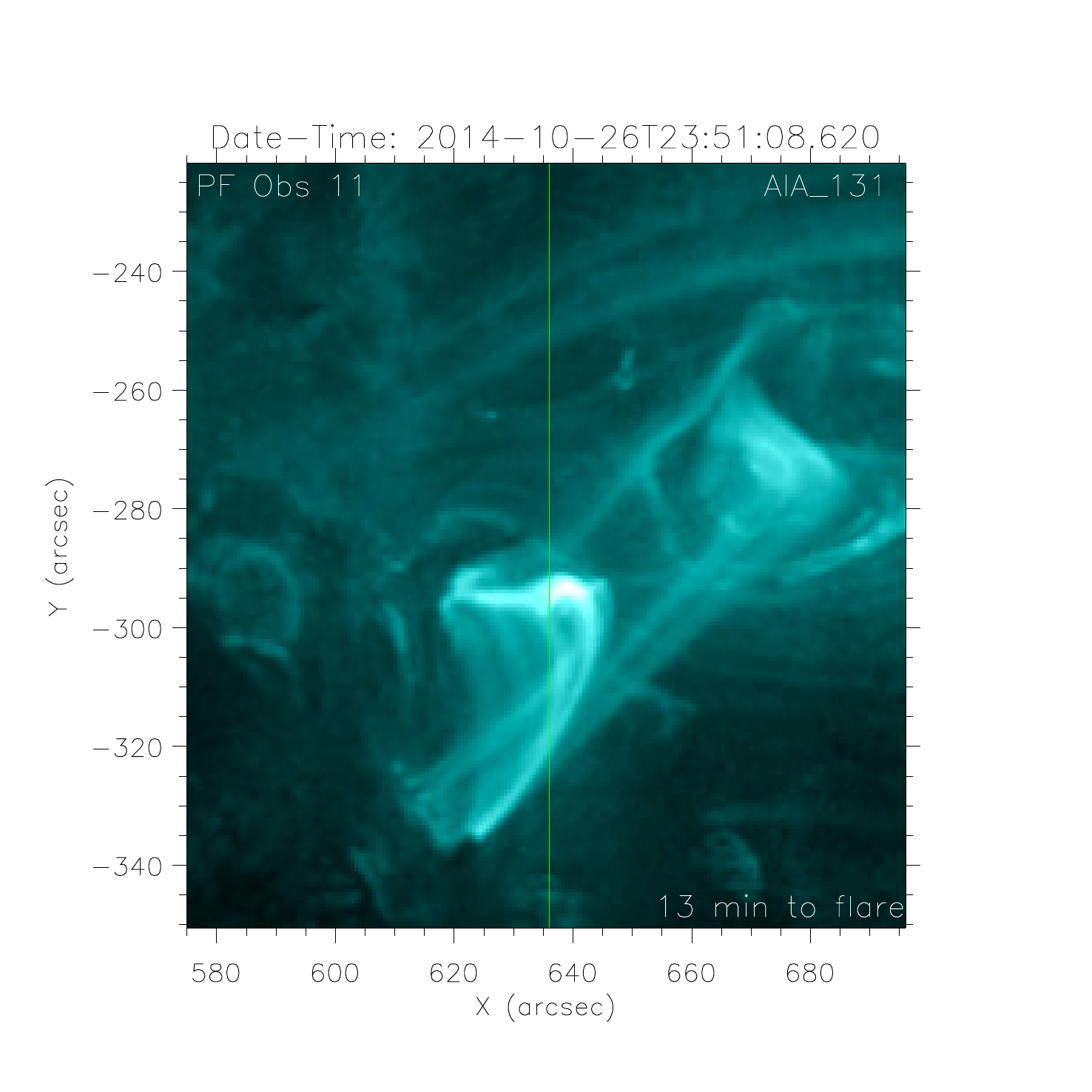}
\caption{Selected images of different PF observations in the strongly predictive regime of Fig~\ref{Integation2}. The PF observation numbering corresponding to Table \ref{obs_table}, as well as the filter wavelength is indicated in the top left and right hand corners of each image, respectively. The green vertical lines indicate the span of the IRIS raster. The flares that are easiest to predict based on spectra have the slit directly over regions where brightenings occur.}
\label{sji_aia}
\end{centering}
\end{figure}

To understand why some flares are easier to predict than others, we analyzed the SJI and AIA movies in every available passband for all PF observations. We found that within the strongly predictive regime, either the span of the rasters or the position of the sit-and stares covered a region that was associated with small brightenings as seen in Fig~\ref{sji_aia}. In contrast, many observations within the weakly predictive regime had poor coverage of the major preflare activity. Additionally, in the high prediction regime, $77\%$ of observations were large rasters ($10$ of $13$ observations) with only 3 sit-and stare observations, this is in contrast to the low prediction regime where only $33\%$ ($2$ of the $6$ observations) were rasters, implying that more spatial coverage of the active region could lead to more reliable increases in prediction score with time. Furthermore, as indicated in the bottom two panels of Fig~\ref{sji_aia}, those sit-and-stares that were in the strongly predictive regimes had the IRIS slit positioned directly over the regions of most preflare activity. Our small sample also appears to indicate that M-class flares are easier to predict than X-class flares, however, this could simply be due to the way these observations were sampled. A notable exception is the X-class flare on March 29, 2014 (PF obs $14$), which despite having excellent coverage from IRIS nevertheless had a prediction curve that decreased before flare onset. The flare was possibly triggered by an erupting filament which showed increased chromospheric Doppler velocities at least an hour before flare onset, as well as plasma heating 15 minutes prior to the filament eruption \citep{Kleint_2015}. A possible explanation for the poor predictive performance, despite the good IRIS raster coverage and physical precursory activity, is that spectra sampled from the filament are out of the models learned distribution, since the filament eruption is rare in the training dataset. Furthermore, the prediction curve might be saturated in the $25$ minutes analyzed and might therefore drop to lower values further away from flare onset.

\section{Conclusions}
We applied a powerful visual network known as a Convolutional Neural Network to IRIS \ion{Mg}{II} spectra with the objective of finding differences between spectral shapes sourced from  preflare regions and active regions that did not lead to a flare. Our model was capable of distinguishing spectra from both these classes with a TSS around 0.8, and a large variance depending on how the training and testing datasets were split.

We then obtained visual explanations using two complementary explainable artificial intelligence techniques called Gradient-weighted Class Activation Mapping and Expected Gradients. These techniques allowed us to automatically discover which features of individual spectra were seen by the model to be most important for the task of flare prediction, representing the highest possible resolution of explanations on the level of the individual wavelengths.

The techniques accomplished this by monitoring the sensitivity of the predictions to either small variations within the network's internal components, or variations over input pixels directly. In both cases the techniques returned similar results up to an additive constant, and allowed us to project heatmaps onto the spectra, with warmer colors (reds, oranges, etc.) indicating more important features than cooler colors (blues) for flare prediction. These heatmaps could then be applied to spectrograms of IRIS in-order to automatically highlight critical regions that could possibly indicate a flare triggering event.

We found that in addition to high triplet emission and core intensity, irregularly shaped profiles, broad spectral cores, and single peaked spectra, flows in the form of extended red and blue wing emissions were also consistently flagged by the model as important precursory features, and that $78$\% of the time, high attribution scores along the IRIS slit were predictive of the location of the flare's maximum UV emission. 

The possible importance of increased turbulence as an early warning sign and potential flare triggering mechanism has been noted in the literature \citep{Harra_2001}, with an example of a rise in nonthermal velocity taking place $11$ minutes before flare onset, possibly stimulated by rising flux. Furthermore, even though high scoring regions are typically associated with enhanced intensities, there are several instances where prediction scores increase while both intensity and the GOES-curve remain constant, once again consistent with the findings of \citet{Harra_2001}, and indicating the importance of spectral analysis for flare prediction at a high resolution.

We note that any practical deployment of a flare prediction strategy for small field of view spacecrafts like IRIS with multiple programmable observational settings (exposure time, cadence, raster mode etc.) is unlikely. The reasons for this are as follows: If we use the prediction score (number of spectra scored confidently by the network as PF), as a "flare warning signal," then the absolute value of the score becomes meaningless due to arbitrarily variations in active region coverage, which depends on the size of the active region itself, as well as programmatic decisions such as IRIS's slit length and area spanned by the raster. All these free variables can either artificially drive the prediction scores up or down. Additionally, features that were important for the binary classification task of separating AR/PF spectra, are not necessarily equally important for telling us when a flare will occur in time. All these considerations imply that the actual objective for IRIS spectral based flare prediction, is not to achieve high TSS scores, but to identify monotonically increasing predictions that have steep inclinations, while disregarding the absolute value of these curves.

We found that on aggregate, the prediction curves increased for the majority of our PF observations closer to flare onset, however, there were several exceptions that did not show any coherent increase in prediction score, and some observations whose scores actually decreased in time. An extensive analysis of each PF observation over all SJI and AIA filters revealed that flares were easier to predict when IRIS sampled a large spatial region or had its slit positioned directly over small brightenings.

Although practical and reliable flare prediction with IRIS is unlikely, the methods developed here could easily be exported to new instruments that can extract spectroscopic data from entire active regions with multi-slit girds, such as NASA's new MUSE mission that has a multi, $37$ slit EUV coronal spectrograph \citep{DePontieu2020}.\\

\begin{acknowledgements}
All ConvNets and XAI techniques were programmed using the deep learning library of PyTorch \citep{PyTorch} and TensorFlow \citep{tensorflow2015}. For the implementation of the $k$-means algorithm, we used the Scikit-Learn module \citep{SK}. The pre-processing was accomplished with IRISreader, a library specifically developed for handling large volumes of IRIS data \citep{IRISreader}. This work was supported by a SNSF PRIMA grant. We are grateful to LMSAL for allowing us to download the IRIS database. IRIS is a NASA small explorer mission developed and operated by LMSAL with mission operations executed at NASA Ames Research Center and major contributions to downlink communications funded by ESA and the Norwegian Space Centre.
\end{acknowledgements}

\bibliographystyle{apj}
\bibliography{journals,references}

\begin{appendix}

\section{Classical neural network}
\label{Classical_neural_network}
\begin{figure}[htb]
\center
\includegraphics[width=0.45\textwidth]{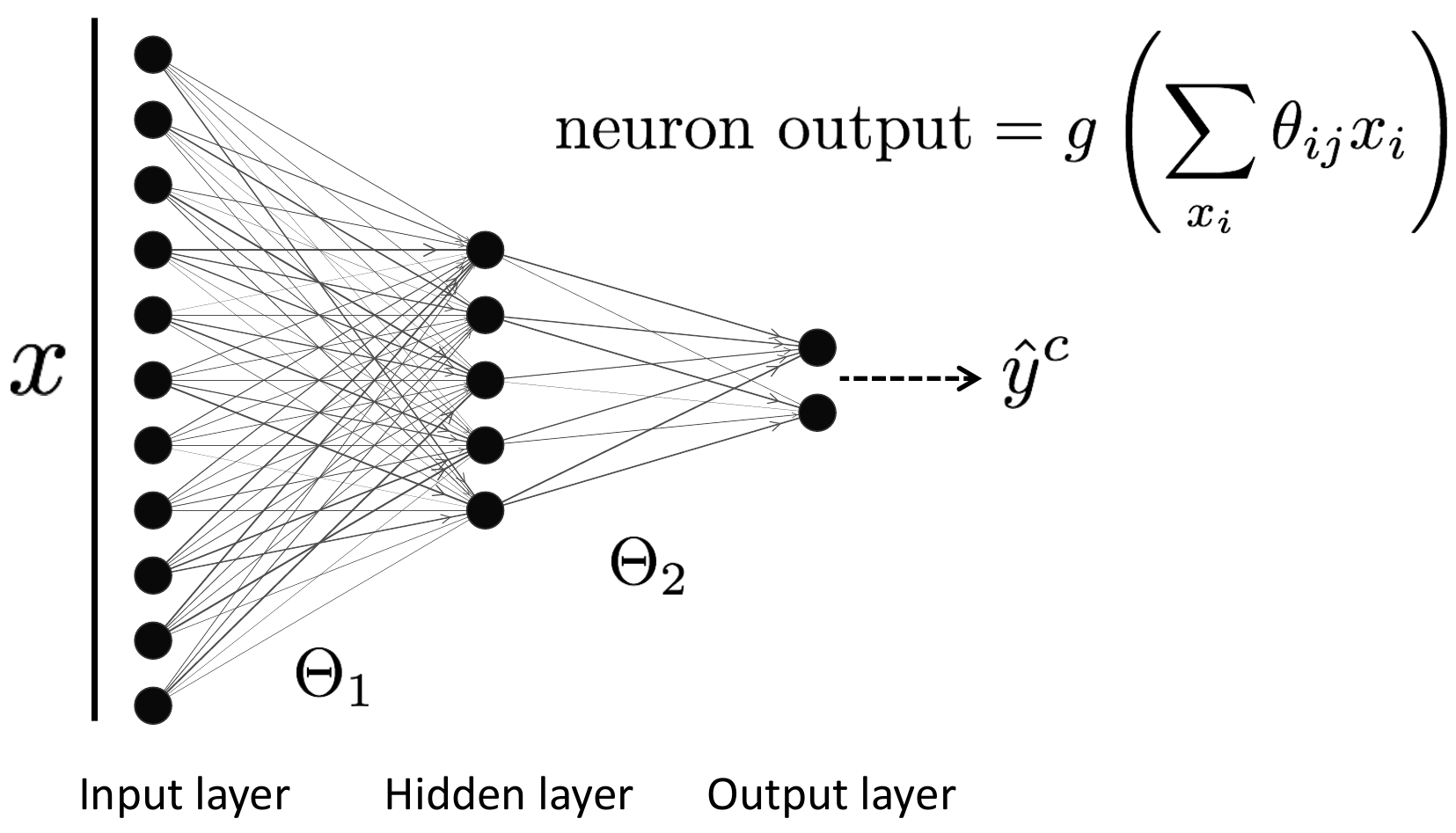} 
\caption{Schematic of a simple fully connected feed forward network. An input spectrum $x$ is fed into the network which stimulates the hidden layer causing the neurons (black circles) to produce output signals that are proportional to the weighted linear sum of the spectrum's intensities. The content of each neuron is then passed through a nonlinear activation function $g$, such as a softmax or ReLU, which allows the network to approximate a large set of complex behaviors. The process is continued with the hidden layer serving as the input layer for the final two neurons. The neuron which produces the strongest signal represents the network's prediction on whether the spectrum comes from the AR or PF class. The weight matrices $\Theta_{1,2}$ determine how information flows through the network and are updated by propagating the errors from a set of predictions back through the network, and adjusting the individual weights against a loss function via gradient descent.}
\label{FFNN}
\end{figure}

One of the simplest foundational networks is the fully connected feed forward network (FNN) shown in Fig~\ref{FFNN}. It consists of multiple layers, with each layer containing a number of neurons represented by black circles.
The term "fully connected" refers to the fact that every neuron has a connection to every other neuron in the layers surrounding it. Each neuron is therefore in communication with all the neurons from the preceding layer, and the strength of the signal between any two neurons is modulated by the models weights $\{\theta\}$. The connections between each layer are therefore summarized by two matrices $\Theta_1$ and $\Theta_2$, where each entry $\theta_{ij}$ from one of these matrices corresponds to the ease of communication between two inter-layer neurons.

The term "feed forward" refers to the fact that information flows forward through the network. The magnitude of the response in neuron $j$ is simply a weighted linear sum of the inputs according to their weights $z=\sum_i\theta_{ij}x_i$. This sum is then passed through a function $g(z)$, referred to as an activation function, which is necessarily nonlinear and allows the network to approximate a large set of complex functions. In our case, the content of the collective stimulation in each of our neurons passes through a ReLU activation function $g(z_j)=z_j^{+}=\max (0, z_j)$ except for the final output layer, which produces a probability by passing the signal through a softmax function $g(z_j)=\exp(z_j)/\sum_i\exp(z_i)$. The two final neurons vote for either the AR or PF class, and the neuron that has the strongest signal represents the network's prediction $\hat{y}^c$.

The randomly initiated weights of the network are optimized by minimizing Eq.~\ref{BCE}. Although prediction is performed by allowing information to flow forward through the network, optimization is performed by propagating the collective errors from the prediction back through the network using an algorithm referred to as backpropagation, a NN equivalent to gradient descent. The network therefore combs through $64$ spectra producing a prediction for each instance. The loss is then calculated using Eq.~\ref{BCE}, and the network's weights are updated using backpropagation. This procedure is continued for several epochs (passes over the entire dataset), adjusting the weights and flow of information through the network, so that some signals are strengthened while others attenuated. 

\begin{figure*}[htb]
\begin{centering}
\includegraphics[width=0.8\textwidth]{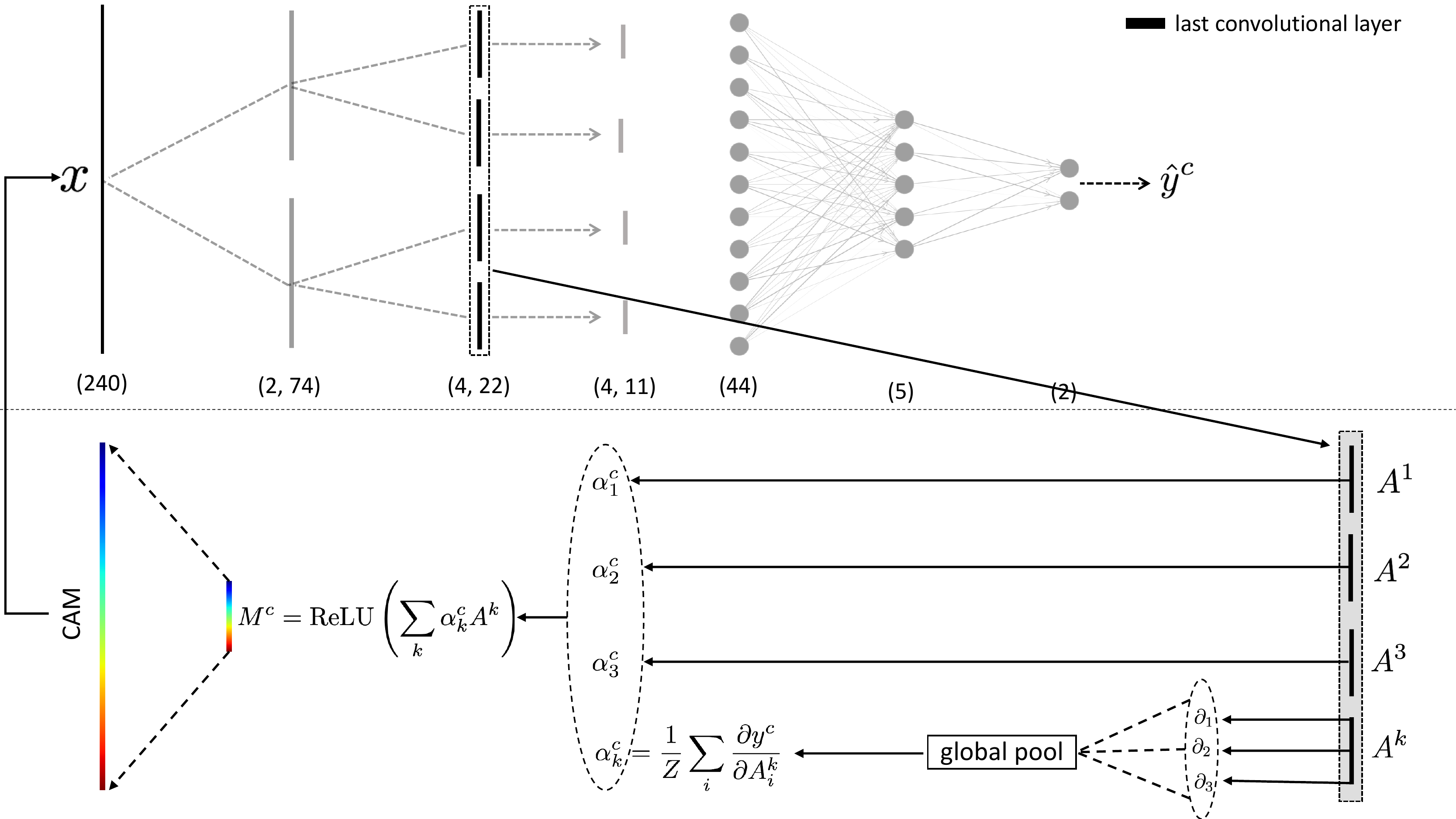}
\caption{Schematic of Grad-CAM. A single spectrum $x$ is passed into a ConvNet that has been trained to distinguish between the PF/AR datasets. The last convolutional layer before the max pooling and fully connected layers contains the network's most sophisticated understanding of the spectrum, with each of the four feature maps $A^k$, focusing on different aspects deemed important for the classification task. The average gradients $\alpha^c_k$ of the output prediction are taken with respect to each of the maps. Larger gradients indicate that the features identified by a particular map are more important for the classification. All four feature maps are then scaled by their corresponding importance and summed before being passed through a ReLU function that isolates the features the network focuses on for the positive classification. The resultant CAM heat map is then interpolated from the scale of the feature map to the dimensionality of the input data before being projected onto the spectrum.}
\label{Grad_CAM}
\end{centering}
\end{figure*}

\section{Grad-CAM (detailed description)}
\label{GradCAMsection_detailed}
We know that each feature map is activated by some visual pattern within the input, such as triplet emission or broad line cores, and furthermore that feature maps in deeper layers are capable of extracting more sophisticated representations of the input \citep{Zeiler_2014}. An important observation when looking at Fig.~\ref{feature_maps}, is to note that the original input spectrum can be roughly reconstructed by placing each of the final feature map activations on-top of one another (summing them together). To makes this clear, feature map $A^1$ identifies the triplet emission, $A^2$ the h-core, $A^3$ the pseudo continuum, and $A^4$ the k-core, thus superimposing all the maps leads to a reconstruction of the original spectrum. Grad-CAM is used to identify which of the four final maps are most important for the classification task, which for our problem translates to which features of each spectrum are strongly related to preflares. After ranking the maps in order of importance, each map can then be assigned a weight before they are summed together. Because each map is weighted differently, the sum of the maps now no longer reproduce the original spectrum, but rather a vector of values whose magnitude at each index represents that pixel's/wavelength's importance. This vector in other words is our heatmap that can be projected back onto the original spectrum.

The question now is how does Grad-CAM assign weights to each map? The mechanism for doing this is depicted in Fig.~\ref{Grad_CAM}. Grad-CAM looks at the maps in the final convolutional layer (dashed box in the figure), and assigns weights to each map by monitoring how sensitive (to first order) the resultant output prediction $y^c$ is to small variations of the map $A^k_i$, where the index $i$ runs over the individual pixels of the map and $c$ represents the class. The weight $\alpha^c_k$ assigned to each map is simply the average (global pooling) of the gradients within each pixel of the map 
\begin{equation}
\alpha^c_k = \frac{1}{Z}\sum_i\frac{\partial y^c}{\partial A^k_i},
\end{equation}
where in our case $Z=22$, the number of pixels in an individual map. The saliency map is therefore given by
\begin{equation}
M^c = \text{ReLU}\left(\sum_k \alpha^c_k A^k\right),
\end{equation}
where the role of the ReLU function is to suppress the contributions from the negative (AR) class, thus producing a saliency map that only highlights the features deemed important for the classification of PF spectra. As seen in Fig.~\ref{Grad_CAM}, the resulting coarse heatmap $M^c$ has to first be interpolated to the size of the input spectrum ($240$ pixels) before it can be projected onto the input and interpreted.

There are several nuances that we list here to clear any confusion about the results produced by Grad-CAM. 1) Although the filters/kernels are fixed in a trained model, different spectra can lead to different activations and therefore different feature maps. This property is referred to as locality, meaning that Grad-CAM produces only local explanations specific to each example spectrum. 2) A single feature map need not find patterns that are spatially connected in the input. For instance, if a kernel learned weights that lead to activations and a feature map that identifies noise, then assuming that the cores have the best signal-to-noise, the maps would disjointly activate over the continuum left and right of both cores. 3) Due to the successive use of convolutions, the feature maps shrink in width as the information flows deeper into the network, meaning that the derived saliency map is often much smaller in size than the input spectrum and has to be interpolated back to the original size. Therefore, network architecture dictates the resolution of the final heatmap, with extremely deep networks having poor heatmap resolution, and extremely shallow networks lacking sophisticated explanations. This last point highlights a weakness of Grad-CAM, namely that the success of quality explanations depends strongly on the selection criterion of the network's architecture, which in itself is done via trial and error.

Our solution to this final point was to construct several network architectures with different depths, number of convolutions, as well as differing structures of the fully connected layers. We found that the problem was simple enough to address with the relatively minimalistic architecture shown in Fig~\ref{Grad_CAM}, and that going deeper resulted in poor resolution, while a single convolutional layer produced poor quality explanations. An ultimate confirmation of our architectural choice however is best supported by using an alternative method that does not itself depend strongly on the specific architecture of the network.

\section{EG (detailed description)}
\label{EG_section_detailed}
EG utilizes the idea of "missingness" which is a common concept in cooperative game theory. The idea is that if we remove wavelengths from the input spectrum, then those missing wavelengths that affect the prediction of the network most must be more critical for its discrimination task. In order to weigh wavelength importance via this method requires one to average over all possible subsets (or coalitions) of wavelengths to ensure the fair distribution of importance.

The need for subsetting becomes clear if one considers the very likely scenario where two neighboring wavelengths contain the same information, and furthermore, that this information is responsible for 90\% of the network's decision. By simply removing one wavelength at a time, the performance of the network will never drop significantly, since the partner wavelength will always carry the discriminant load. Counter-intuitively, this might result in the two most important wavelengths being ranked as useless, which is not such an unlikely scenario if one considers that proximal wavelengths in a high-resolution spectrogram such as IRIS are likely to transmit similar types of information.

A complete formalism that allows for the fair distribution of wavelength relevance can be found in the game theoretic quantity called the Shapley value \citep{Shapley1951}, given by
\begin{equation}
\phi_{\lambda}(\mathcal{F}_\Theta)=\sum_{S \subset N} \frac{(s-1) !(n-\mathrm{s}) !}{n !}[\mathcal{F}_\Theta(S)-\mathcal{F}_\Theta(S-\lambda)],
\label{shapley_eqn}
\end{equation}
where $|N|=n=240$ is the total number of $\lambda$-points, $|S|=s$ is a subset of those wavelengths, $\lambda$ represents a single input wavelength, whose importance we are trying to calculate, and $\mathcal{F}_\Theta$ is the characteristic function giving us the subset's "worth". If we parameterize the characteristic function $\mathcal{F}_\Theta$ with a ConvNet trained on the binary classification PF/AR problem, then we can interpret the Shapley value for a particular wavelength $\lambda$ and input spectrum $x$, as the expected difference between the prediction $\hat{y}$ over all lambda coalitions/subsets, with and without said wavelength. 

The prefactor $(s-1)!(n-s)!$/n! weights each score discrepancy by the number of ways a particular subset of pixels can be formed (in our case each of the n-wavelengths are equally likely), and allows us to alternatively interpret the Shapley value $\phi_{\lambda}$ as the expected marginal contribution of $\lambda$ to the network's output prediction.\\

The question now is how to reformulate Eq.~\ref{shapley_eqn} so that it is compatible with functions $\mathcal{F}_\Theta$ that are parameterized by NNs. An important caveat is that NNs have fixed architectures, so one cannot simply remove input channels when subsetting over wavelength coalitions. This means that NNs require each wavelength to pump information through it at all times. A solution is therefore to construct a baseline that represents missingness and can be fed through the network constantly without affecting the results. In this way, we can artificially turn some wavelengths off while still respecting the constant flow of information through the networks input channels, allowing us to form pseudo subsets to approximate the form of Eq.~\ref{shapley_eqn}. 

This baseline must be selected with great care, since it can severely affect and bias the results. For instance, if one assumes that missingness is best represented by zeros, then the black regions of any image (such as outline contours) or low intensity portions of any spectrum (which might be the most critical aspects) would be considered useless. Another common choice of baseline is to inject Gaussian noise into some wavelengths so that with many forward passes through the network, the average contribution of said wavelength becomes negligible \citep{Smilkov_2017}. This baseline however ignores the very real fact that neighboring, as well as disjoint wavelengths might be correlated to one another, and not independently distributed as such distributions would assume. 

Arguably, the correct baseline, and the strategy employed by EG, is to represent missingness by the actual dataset itself. Like the Gaussian distribution, flooding combinations/subsets of wavelengths with the intensities found within the dataset, would result in a zero net average displacement of the network's prediction score for a particular spectrum. To see why, consider that after randomly sampling particular combinations of intensities from the dataset, it is likely that some features emerge from the AR class, leading to a decrease in the network's output prediction. At the same time, the sampling is equally likely to counterbalance this shift by allowing particular features associated with the PF class to emerge, resulting in an increase to the score. In addition to the invariant behaviour of the score under this type of sampling (a fact that is critical for the formation of pseudo subsets), it also respects the full set of correlations and inter-dependencies that exist between the wavelengths.
\begin{figure}[thb]
\center
\includegraphics[width=0.48\textwidth]{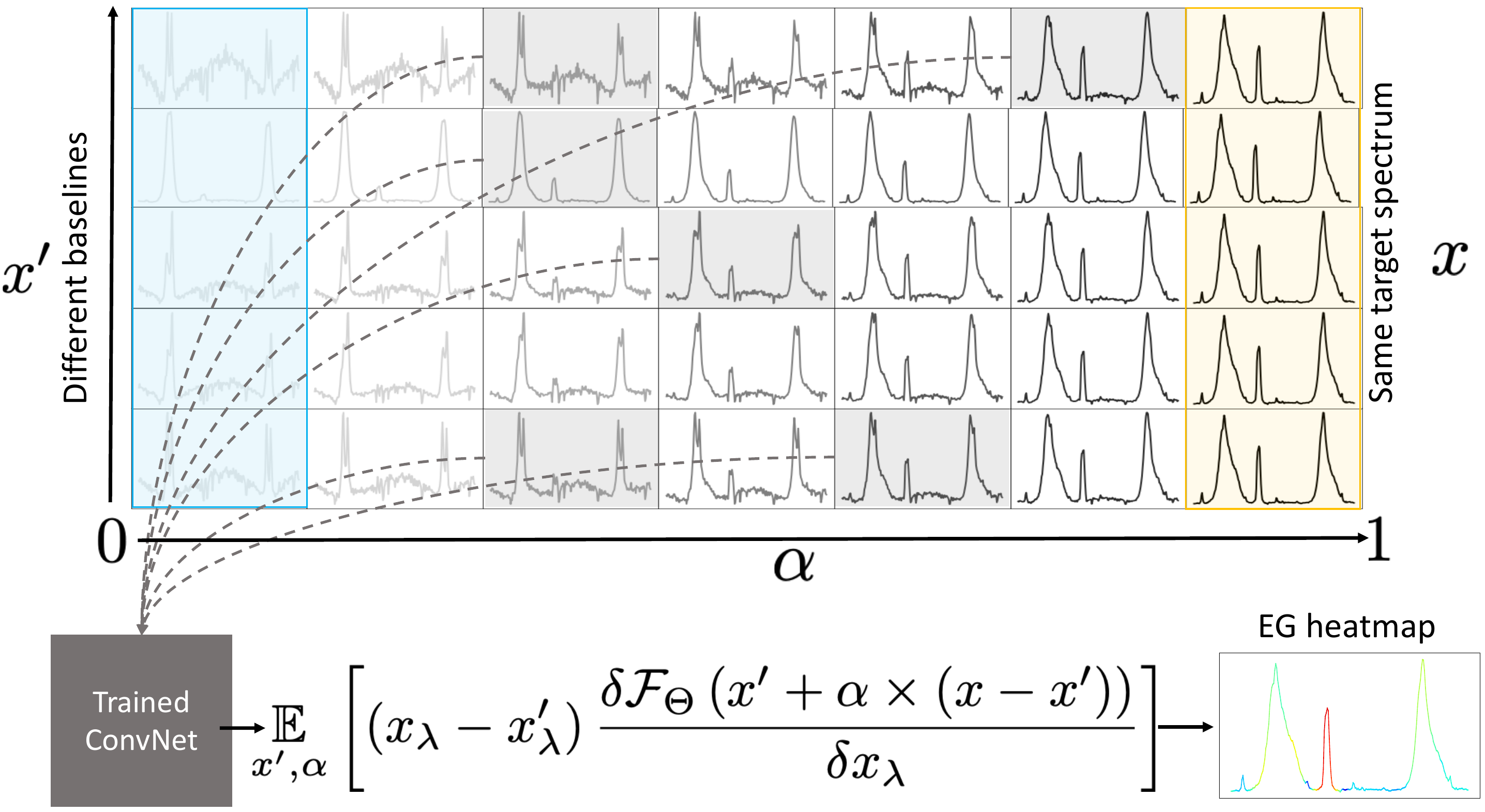}
\caption{Diagram showing the calculation procedure for producing an EG heatmap. The heatmap for a target spectrum (orange) is calculated by interpolating between a set of baseline spectra (blue) and the target. As we move from left to right along each row, the interpolated spectrum in each panel is fed into the network which generates a prediction score. The wavelengths with the largest accumulated gradients along the path are then ranked as most critical for flare prediction. Instead of integrating over all paths, EG calculates an expectation value for the wavelength's importance by feeding a random sample (gray spectra) into the network.}
\label{EG_path}
\end{figure}

Figure~\ref{EG_path} demonstrates how EG derives the heatmap for a particular PF spectrum. Each of the rows in the figure represent a path in image space. In each case, we start from some baseline for missingness, that is, a specific spectrum from the conjoined PF/AR dataset (blue). In each row, EG interpolates from left to right between the baseline spectrum and a particular target spectrum (orange). Notice that although the baseline starts from different points in the image space, that is, different spectra, the target is always the same spectrum for which we want to calculate the heatmap. As we vary $\alpha$ from $0$ to $1$, the features of the target spectrum (such as triplet emission) slowly emerge from the baseline and become dominant. At each point along a particular path, we feed the interpolated spectra into the network $\mathcal{F}_\Theta$ and calculate its prediction score.

We can then estimate which wavelengths contributed most to the output score by calculating the difference between scores of successive images along a path and monitoring the accumulated gradient of the output with respect to each wavelength. Those wavelengths that are critical for the network's decision will affect the output's prediction most. For instance, in the case shown here, the triplet emission is not present at the start of the path, but slowly emerges as we get closer to the target. If the network thinks this feature is important, the output will depend most strongly on these wavelengths and accumulate the steepest gradients as we travel along the image path. 

We therefore understand that integrating along $\alpha$ while monitoring the accumulated gradients along a single path is analogous to calculating the difference between scores under a single subset, reproducing the functionality of $[\mathcal{F}_\Theta(S)-\mathcal{F}_\Theta(S-\lambda)]$ in Eq.~\ref{shapley_eqn}. However this subset at the moment is "dirty", since we have a single baseline from the dataset whose effects only become negligible when averaged over many paths. In order to unbiase our baseline, as well as form pseudo subsets over every combination of wavelengths, we need to gain an additional integration term over $x^\prime$ and integrate over all image paths. 

Putting this all together, a mathematically concise NN equivalent of Eq.~\ref{shapley_eqn} is given by
\begin{equation}
\begin{aligned}
\phi_\lambda|_\text{EG} &=\int_{x^{\prime}}\Bigl(\left(x_\lambda-x_\lambda^{\prime}\right) \times
\\ & \times \int_{\alpha=0}^{1} \frac{\delta \mathcal{F}_\Theta \left(x^{\prime}+\alpha\left(x-x^{\prime}\right)\right)}{\delta x_\lambda} d \alpha\Bigr) p_{D}\left(x^{\prime}\right) d x^{\prime}
\\ &\simeq\underset{x^{\prime} \sim D, \alpha \sim U(0,1)}{\mathbb{E}}\left[\left(x_\lambda-x_\lambda^{\prime}\right) \frac{\delta \mathcal{F}_\Theta\left(x^{\prime}+\alpha \times\left(x-x^{\prime}\right)\right)}{\delta x_\lambda}\right],
\label{EG_eqn}
\end{aligned}
\end{equation}
where the first integral over $x^\prime$ selects the baseline starting point, and at heart is responsible for generating our pseudo subsets over wavelength allegiances. The second integral over $\alpha$ propagates us along a particular path as defined by $x^\prime$, such that the target spectrum (orange) emerges from the baseline as $\alpha|0\to1$, allowing us to calculate the differences between scores with and without particular features. The function $\mathcal{F}_\Theta$ as usual represents our ConvNet, which is the same as that used for Grad-CAM, and was trained on the binary PF/AR classification problem. Finally $D$ means that our baseline samples form the actual dataset, which as discussed above results in a net zero effect on the output score when averaged over many samples. Crucially, this choice of baseline also respect the existence of a rich set of correlations that exist between different wavelengths, something a Gaussian analog for missingness would neglect.

Since the process of integrating over all complete image paths is computationally infeasible, EG takes the expectation value along a set of randomly selected paths. In Fig.~\ref{EG_path} this is represented by only feeding the NN the spectra from the randomly selected gray highlighted panels. It has been shown that the expectation converges and that the results respect several important interpretability axioms such as completeness, where each wavelength's attributions sums to the network's output prediction \citep{Erion_2021}. For our purposes we found that a baseline sample of $500$ spectra was sufficient for EG to converge.

Normally the heatmaps derived from such path attribution methods do not necessarily produce smooth outputs, that is, attributions between neighboring wavelengths can be large. In many cases one would assume that neighboring wavelengths would relay similar types of information and therefore produce smooth attributions. Much work has gone into encouraging such intuitive restrictions over the derived explanations \citep{Rieger_2019}. For EG, this is normally achieved by incorporating the attributions into a prior, and wrapping them into a differentiable function that promotes smooth attributions across neighboring wavelengths, while penalizing heatmaps with large amounts of local variation. In this way, the attributions form part of the training procedure, and on some datasets actually improve model performance and convergence time \citep{Erion_2021}. We attempted to incorporate this prescription by defining the generalized loss function
\begin{equation}
\mathcal{S}(\Theta, x) = \underbrace{\mathcal{L}(\Theta, X)}_\text{BCE} + \underbrace{\sigma \sum_{x,\lambda}| \phi(x)_{\lambda+1} - \phi(x)_\lambda |}_\text{attribution prior},
\end{equation}
with the first term $\mathcal{L}$ being the standard binary cross-entropy as discussed in Eq.~\ref{BCE}, and the second term a differentiable attribution prior consisting of wavelength attributions $\phi(x)_\lambda$ from Eq.~\ref{EG_eqn}. This term promotes the variations between attributions of neighboring wavelengths to be small. Here, $\sigma$ is a regularization parameter that controls the trade-off between smooth attributions and model accuracy. 

Unfortunately, following this prescription did not result in smooth heatmaps for the case of our data. The smoothing effect only initiated after the classification score became small, which tended to always be after the point where the model overfitted the training data, resulting in a divergent validation loss. When we tried to increase the hyperparameter $\sigma$ such that the smoothing would initiate before the critical overfitting point, the network's classification score suffered and rapidly diverged. We therefore selected to add a manual small smoothing post-hoc (after training) for increased interpretability. The degree of smoothing was adjusted until the heatmaps matched those derived from grad-CAM.

As a point of analogy, grad-CAM naturally smooths its attributions due to the coarse feature maps that result from the dimensionality bottleneck brought about by consecutive convolutional layers. In this case attributions are assigned on an $22$-point grid (final feature map dimension) and then upscaled to the resolution of the input. EG on the other hand always operates and derives its attributions at the resolution of the input.

\section{Supplementary figures}

\begin{figure}[htb]
\begin{centering}
\includegraphics[width=0.5\textwidth]{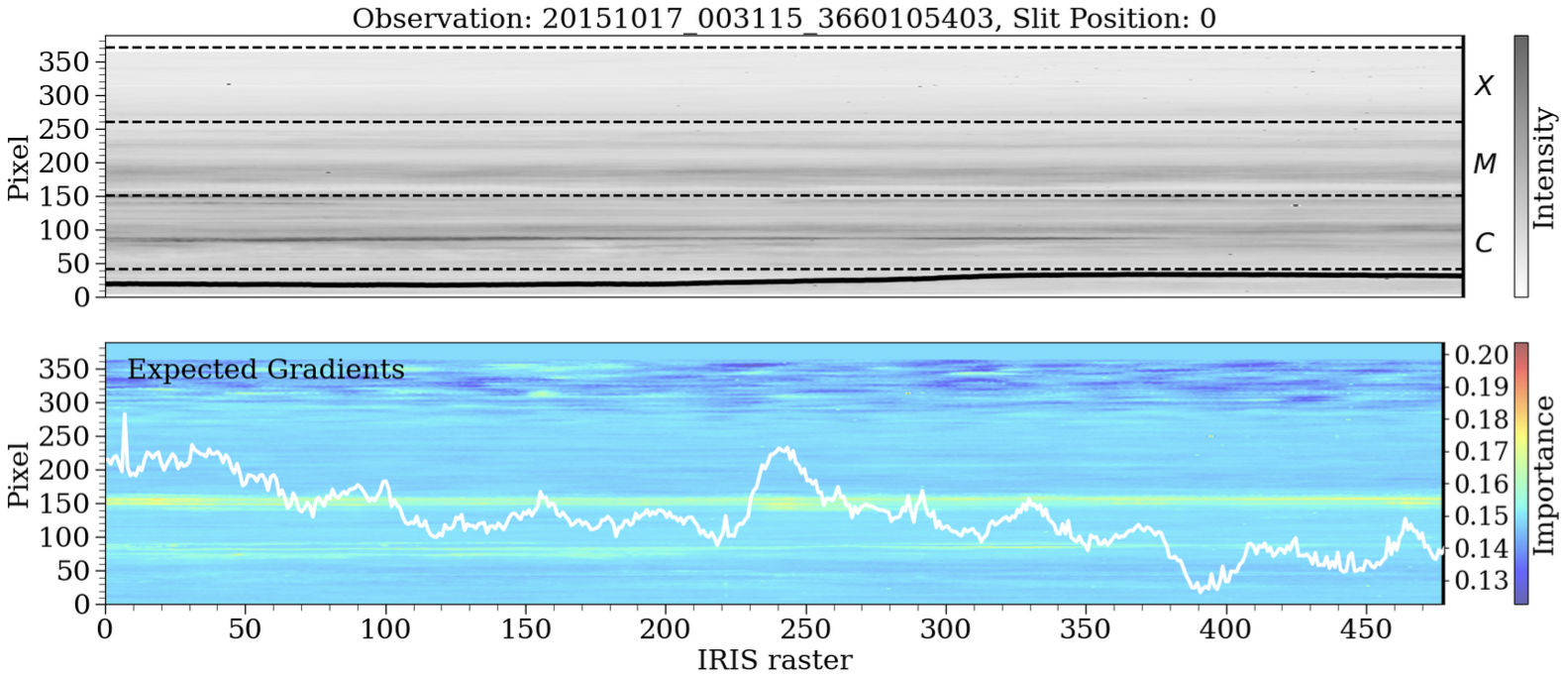}
\includegraphics[width=0.5\textwidth]{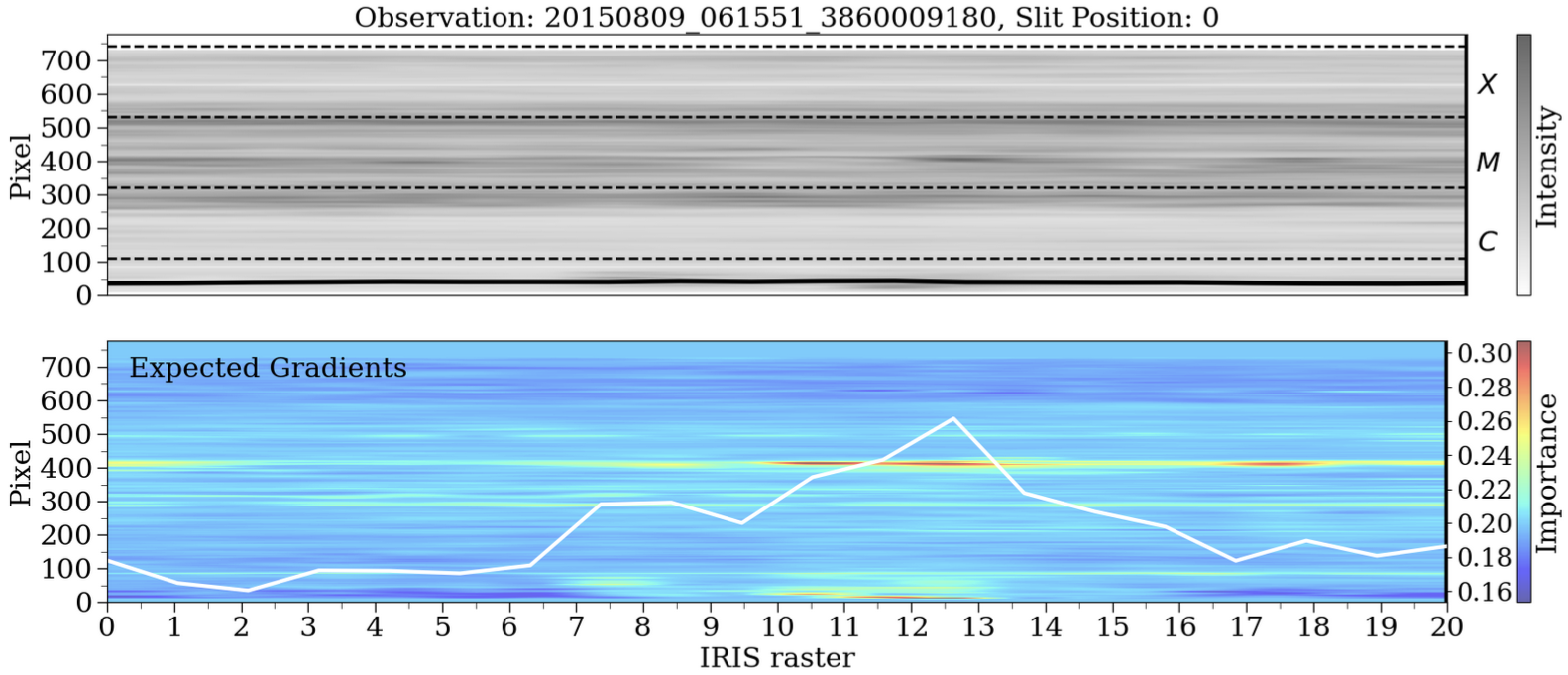}
\caption{Same as Fig.~\ref{Integration1}, but for nonflare (AR) events.}
\label{Integration_neg}
\end{centering}
\end{figure}

\begin{figure}[htb]
\begin{centering}
\includegraphics[width=0.5\textwidth]{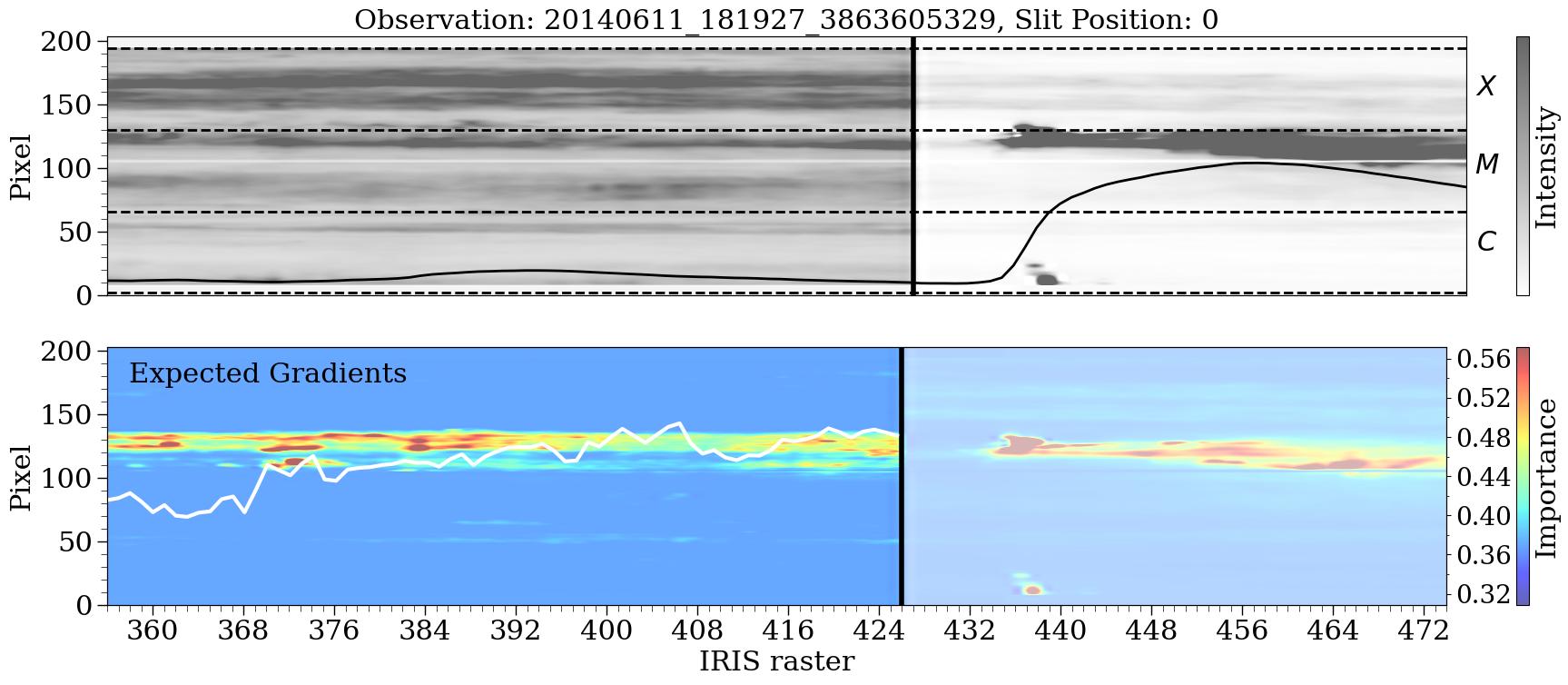}
\caption{Same as Fig.~\ref{Integration1}, but an example of how the location of the maximum attributions along the slit in the preflare region (bottom panel, left) align with the maximum UV emission of the flare (upper panel, right). Notice that even though the intensity is larger around pixel $160$, the attributions are only high around pixels $130$, where the maximum UV intensity is seen later during the flare.}
\label{aligned_pic}
\end{centering}
\end{figure}

\setcounter{table}{0}
\renewcommand{\thetable}{A\arabic{table}}

\begin{table*}[ht]
\caption{PF observations\label{obs_table}}
\begin{tabular*}{\textwidth}{lllllll}
\toprule\toprule
\# ~~~&~~~~~~Class  &  ~~~~~~Date and time &  ~~~~~~Observation mode   & ~~~~~~CAD   &~~~~~~FOV center & ~~~~~~OBSID \\
& & ~~~~~~when raster started& & ~~~~~~(sec) &~~~~~~(arcsec)&\\\midrule
 1& M6.5&2015-06-22T17:00&Large sparse 16-step raster&33 &(72,192)&3660100039\\
 2& M1.0&2014-11-07T09:37& Large coarse 16-step raster& 23 & (-646,224)&  3860602088\\
 3& M1.1&2015-08-21T16:01& Medium dense 32-step raster& 102 & (-467,-336)& 3660104044\\
 4& M3.9&2014-06-11T18:19&Medium coarse 8-step raster&21&(-781,-306)&3863605329\\
 5& M1.8&2015-03-11T04:46&Large coarse 8-step raster&75 &(-430,-194)&3860259280\\
 6& M1.1& 2014-09-06T11:23& Large sit-and-stare& 9 & (-709,-298)& 3820259253 \\
 7& M3.4&2014-10-27T20:56&Large sit-and-stare&16 &(779,-271)&3864111353\\
 8& X2.1&2015-03-11T15:19&Large coarse 4-step raster&16&(-353,-197)&3860107071\\
 9& M1.0&  2014-06-12T18:44& Medium coarse 8-step raster&  21 & (-670,-306)& 3863605329\\ 
 10& M1.8&2014-02-12T21:50&Large coarse 8-step raster&42 &(140,-90)&3860257280\\
 11& M1.3& 2014-10-26T18:52& Large sit-and-stare& 16 & (648,-287)& 3864111353\\
 12& X2.0&2014-10-27T14:04&Large coarse 8-step raster&26 &(727,-299)&3860354980\\
 13& M2.3&2014-11-09T15:17&Large coarse 4-step raster&37 &(-217,-205)&3860258971\\
 14& X1.0&2014-03-29T14:09&Very large coarse 8-step raster&72 &(490,282)&3860258481\\
 15& X1.6&2014-10-22T08:18&Very large coarse 8-step raster&131 &(-292,-303)&3860261381\\
 16& M1.4& 2015-03-12T05:45& Large sit-and-stare& 5 & (-185,-190)& 3860107053\\
 17& X1.0&2014-10-25T14:58&Large sit-and-stare&5 &(408,-319)&3880106953\\
 18& X1.6&2014-09-10T11:28&Large sit-and-stare&9&(-137,125)&3860259453\\
 19& M8.7&2014-10-21T18:10&Large sit-and-stare&16 &(-359,-316)&3860261353\\
\midrule
\midrule
&&&~~~~~~~~~~~AR-observations&&\\
\midrule
\textcolor{black}{1} &&  \textcolor{black}{2015-05-18T14:39} & \textcolor{black}{Large coarse 4-step raster} & ~~\textcolor{black}{21}  & \textcolor{black}{(300,-98)} & \textcolor{black}{3860256971} \\ 
2 &&  2015-05-18T16:14 & Large coarse 4-step raster & ~~21  & (315,-95) & 3860256971 \\ 
3 &&  2015-05-21T18:59 & Very large sit-and-stare & ~~5  & (-382,398) & 3800507454 \\ 
4 &&  2015-07-03T16:59 & Large sparse 8-step raster & ~~45  & (-186,213) & 3620006130 \\ 
5 &&  2015-07-04T10:09 & Very large sit-and-stare & ~~9  & (86,174) & 3860108354 \\ 
6 &&  2015-07-04T16:59 & Large sparse 8-step raster & ~~44  & (20,202) & 3620006130 \\ 
7 &&  2015-07-28T15:18 &Medium coarse 4-step raster & ~~37  & (-227,-289) & 3660109122 \\ 
8 &&  2015-08-07T22:14 & Large coarse 8-step raster & ~~74  & (547,125) & 3860259180 \\ 
\textcolor{black}{9} && \textcolor{black}{2015-08-09T06:15} & \textcolor{black}{Large coarse 8-step raster} & ~~\textcolor{black}{75}  & \textcolor{black}{(-236,-370)} & \textcolor{black}{3860009180} \\ 
10 && 2015-09-16T18:17 & Medium coarse 16-step raster & ~~34  & (-564,-356) & 3600101141 \\ 
11 && 2015-10-17T00:31 & Large sit-and-stare & ~~3  & (-558,-233) & 3660105403 \\ 
\textcolor{black}{12} && \textcolor{black}{2015-07-24T05:35} & \textcolor{black}{Large sit-and-stare} & ~~\textcolor{black}{9}  & \textcolor{black}{(557,-204)} & \textcolor{black}{3620109103} \\
\textcolor{black}{13} && \textcolor{black}{2015-04-08T04:57} & \textcolor{black}{Large sit-and-stare} & ~~\textcolor{black}{5} & \textcolor{black}{(45,-118)} & \textcolor{black}{3860107054}\\
14 && 2015-01-30T11:27 & Very large dense 4-step raster & ~~21  & (-756,161) & 3860607366 \\
15 && 2014-03-29T20:14 & Medium sit-and-stare & ~~17  & (687,-166) & 3820011652 \\
16 && 2014-12-01T15:44 & Large sit-and-stare & ~~10  & (-80,-329) & 3800008053 \\
17 && 2014-03-13T09:35 & Large sit-and-stare & ~~9  & (521,23) & 3820109554 \\
18 && 2014-11-28T21:05 & Very large sit-and-stare & ~~10  & (-34,-322) & 3860009154 \\
\bottomrule
\end{tabular*}
\end{table*}

\end{appendix}
\end{document}